\documentclass[11pt]{article}
\usepackage{CJKutf8}
\usepackage[margin=1in]{geometry} 
\usepackage{amsmath,amsfonts}
\usepackage{algorithmic}
\usepackage{algorithm}
\usepackage{graphicx}
\usepackage{array}
\usepackage[caption=false,font=normalsize,labelfont=sf,textfont=sf]{subfig}
\usepackage{textcomp}
\usepackage{stfloats}
\usepackage{url}
\usepackage{verbatim}
\usepackage{cite}
\usepackage{xcolor} 
\usepackage{hyperref}
\usepackage{authblk}  
\usepackage{fvextra} 
\usepackage{multirow}
\usepackage{booktabs}
\usepackage{tabularx}
\usepackage{pifont}    

\newcommand{\cmark}{\ding{51}}  
\newcommand{\na}{--}            

\title{A Multimodal fNIRS-EEG Dataset for Unilateral Limb Motor Imagery}

\author[1]{Lufeng Feng}
\author[1]{Baomin Xu}
\author[1]{Haoran Zhang}
\author[1]{Bihai Lin}
\author[1]{Zuxuan Deng}
\author[1]{Sidi Tao}
\author[2]{Chenyu Liu}
\author[3]{Shifan Jia}
\author[1*]{Li Duan}
\author[4*]{Ziyu Jia}
\affil[1]{Beijing Jiaotong University, Beijing, 100044, China}
\affil[2]{Nanyang Technological University, 50 Nanyang Avenue, Singapore}
\affil[3]{Simon Fraser University, Vancouver V5A 1S6, Canada}
\affil[4]{Institute of Automation, Chinese Academy of Sciences, Beijing, 100044, China}
\affil[*]{Corresponding authors: Li Duan (duanli@bjtu.edu.cn) and Ziyu Jia (jia.ziyu@outlook.com) }
\date{} 
\begin{document}
\begin{CJK*}{UTF8}{gbsn}
\maketitle

\begin{abstract}

Unilateral limb motor imagery (MI) plays an important role in upper-limb motor rehabilitation and precise control of external devices, and places higher demands on spatial resolution. However, most existing public datasets focus on binary- or four-class left–right limb paradigms that mainly exploit coarse hemispheric lateralization, and there is still a lack of multimodal datasets that simultaneously record EEG and fNIRS for unilateral multi-directional MI. To address this gap, we constructed MIND, a public motor imagery fNIRS–EEG dataset based on a four-class directional MI paradigm of the right upper limb. The dataset includes 64-channel EEG recordings (1000~Hz) and 51-channel fNIRS recordings (47.62~Hz) from 30 participants (12 females, 18 males; aged 19.0--25.0~years). We analyse the spatiotemporal characteristics of EEG spectral power and hemodynamic responses, and validate the potential advantages of hybrid fNIRS--EEG BCIs in terms of classification accuracy. We expect that this dataset will facilitate the evaluation and comparison of neuroimaging analysis and decoding methods.
\end{abstract}

\section*{Background \& Summary}
Brain–computer interfaces (BCIs) are systems that acquire and decode brain activity signals and translate them into control commands for external devices~\cite{vaid2015eeg}. According to the signal acquisition modality, BCIs can be divided into invasive and non-invasive types~\cite{he2020brain}. Motor imagery (MI) is one of the most importance paradigms in non-invasive BCIs~\cite{al2021deep}. With the rapid expansion of BCIs into clinical practice, rehabilitation, and human–computer interaction, the potential application scenarios of MI have become increasingly diverse, including upper-limb functional reconstruction~\cite{ang2015randomized, mane2020bci}, daily assistive control~\cite{limchesing2021review} and continuous trajectory~\cite{chen2022continuous} guidance.

\begin{figure}[h!]
	\centering
    \includegraphics[scale=0.45]{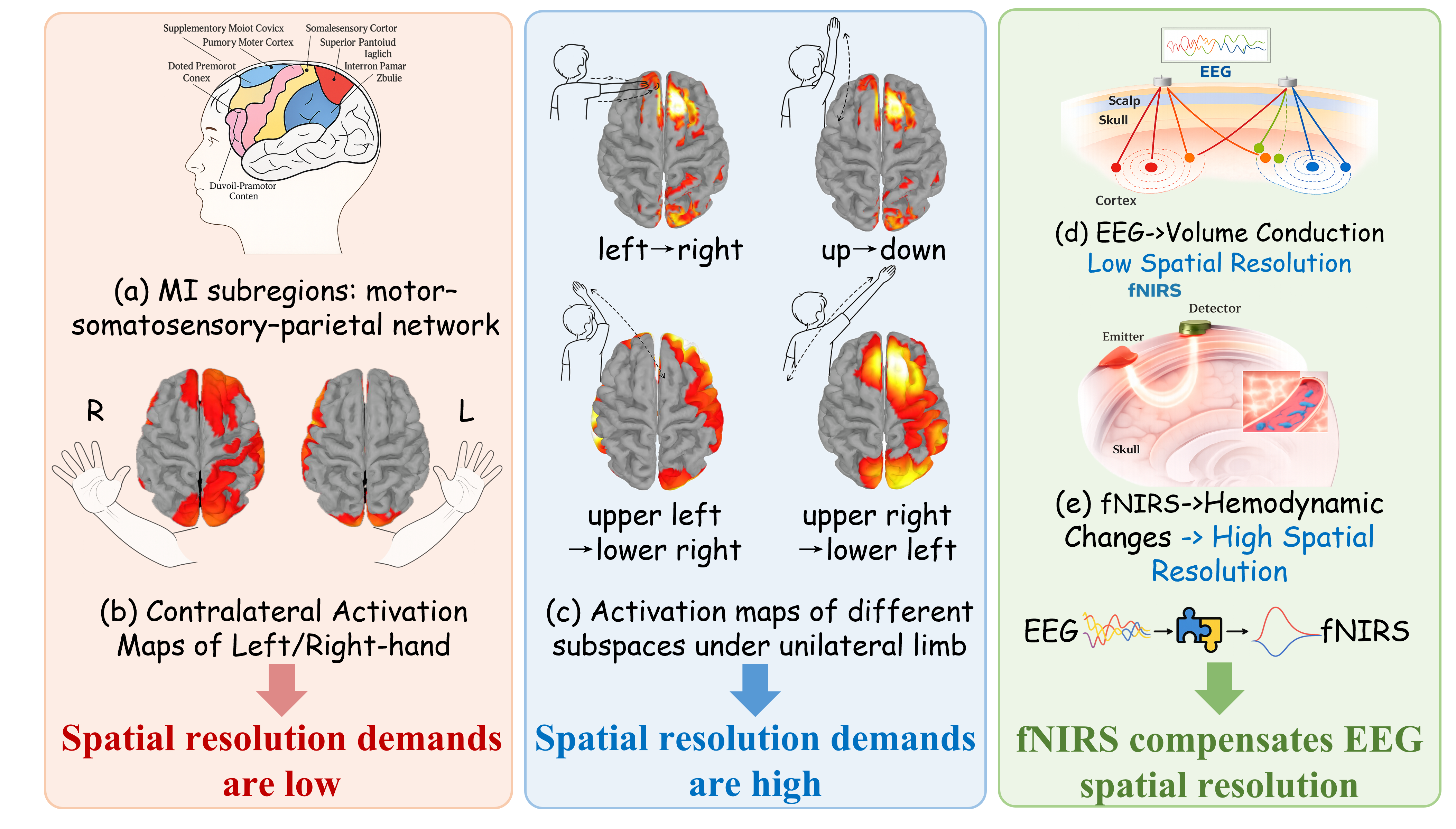}
    \caption{Motivation framework of the dataset. (a) MI subregions: motor–somatosensory–parietal network (M1/S1, PMd/PMv, SMA, and SPL/IPS). (b) Contralateral Activation Maps of Left/Right-hand. (c) Activation maps of different subspaces under unilateral limb. (d) Mechanism of EEG recording. Due to volume conduction, EEG has low spatial resolution. (e) Mechanism of fNIRS recording. fNIRS measures local $\Delta$HbO/$\Delta$HbR with higher spatial specificity. All activation maps are plotted on a common cortical surface; brighter colors indicate stronger MI-related activity.}
	\label{motivation}
\end{figure}

MI involves a distributed motor--somatosensory--parietal network, including the primary motor cortex (M1), primary somatosensory cortex (S1), lateral premotor cortex (PMd/PMv), supplementary motor area (SMA), and the superior parietal lobule / intraparietal sulcus (SPL/IPS) (Fig.~\ref{motivation}(a))~\cite{Graziano2002cotex}. 
Traditional MI paradigms mainly focus on binary- or four-class bilateral limb tasks that are well separated in spatial location (e.g., left vs. right hand, feet and tongue MI)~\cite{al2021deep}. These paradigms are commonly used for algorithm development and benchmarking~\cite{zhang2023classification}. 
In these tasks, class discrimination mainly relies on hemispheric lateralization differences. As shown in Fig.~\ref{motivation}(b), simply identifying ``which hemisphere is more active'' is often sufficient to separate classes, so the requirement for spatial resolution is relatively low~\cite{wang2024upper}. 
However, in application scenarios that require higher-dimensional control commands, such as upper-limb stroke rehabilitation~\cite{wang2024rehabilitation}, unilateral exoskeleton control~\cite{tang20232024exoskeleton}, or prosthetic control~\cite{Ortiz2020arm}. Traditional MI paradigms provide limited classes and control degrees of freedom, and may not meet the need for fine and continuous upper-limb control.

Unilateral MI with multiple directions and multiple joints has a more intuitive correspondence to practical needs(e.g., upper-limb rehabilitation, and daily assistive control). Without substantially increasing cognitive load, unilateral MI can provide higher control freedom and better action separability~\cite{Rong2024limbs, chu2023joint}, and thus has become an active topic in MI-BCI research in recent years~\cite{Yi2025joints}. Unilateral multi-class MI typically does not show strong cross-hemisphere contrasts. Instead, class differences are reflected by subtle spatial modulations within the motor--somatosensory--parietal network of the same hemisphere.~\cite{Pilgramm2016hand}. For example, in M1/PMC/SPL, representations of the hand, wrist, elbow, and shoulder are not isolated, non-overlapping ``blocks''~\cite{Graziano2002cotex}. They form a proximal--distal somatotopic gradient along the precentral gyrus~\cite{Ejaz2015sensorimotor}. Different movement directions (e.g., rightward abduction vs.\ right-upward lifting) often manifest as small shifts of the activation centroid within this map, and as differences in coupling with parietal regions involved in spatial attention and motor planning (e.g., SPL/IPS)~\cite{Andersen2014brain}. Fig.~\ref{motivation}(c) illustrates local cortical differences across four unilateral tasks, suggesting that directional MI differences are finer-grained and thus impose higher demands on imaging and decoding spatial resolution.

Compared with bilateral paradigms, unilateral multi-joint and multi-direction MI depends more on high-precision spatial information. Due to volume conduction, each scalp EEG electrode (e.g., around C3, C1, and CP3) records a weighted mixture of multiple nearby cortical sources (Fig.~\ref{motivation}(d)), leading to low spatial resolution~\cite{he2018esi}. Therefore, relying on scalp EEG alone can cause source mixing, which averages out subtle source-level differences and limits unilateral MI decoding performance. Moreover, most existing unilateral directional multi-class datasets are EEG-only. To fill this gap, we introduce functional near-infrared spectroscopy (fNIRS), which provides higher spatial specificity. fNIRS measures local hemodynamic responses via $\Delta$HbO/$\Delta$HbR, offering a more spatially precise description of cortical activation (Fig.~\ref{motivation}(e)) and complementary spatial information to EEG, as supported by prior studies~\cite{sun2024fnirs}. In addition, fNIRS has relatively good cortical spatial resolution, is more robust to EMG/electrical noise artefacts, and can be recorded simultaneously with EEG without mutual interference. Compared with fMRI, fNIRS is low-cost, portable, and practical,  making it suitable for wider use~\cite{codina2025multimodal}.


To compensate for the limited spatial resolution of EEG, we design a multimodal fNIRS–EEG unilateral upper-limb four-class directional motor imagery paradigm and release an open dataset, named \underline{M}otor \underline{I}magery f\underline{N}IRS--EEG \underline{D}ataset (MIND). The MIND dataset contains EEG and
fNIRS recordings from 30 participants. EEG was recorded from 64 electrodes
positioned according to the international 10--20 system, and fNIRS was
recorded from 51 measurement channels. Each participant completed
three blocks (six sessions in total). Each session included a 60~seconds resting
state and trials of two 10~seconds unilateral upper-limb MI conditions. In total,
each participant performed 120 trials (30 trials per MI class), yielding
3,600 EEG and fNIRS trials across all subjects.

The dataset provides raw recordings to support flexible data analysis and
benchmarking of decoding algorithms. To demonstrate data quality, we
performed preliminary time-, frequency- and topography-domain analyses of
the simultaneously acquired fNIRS--EEG signals. We further used standard
machine-learning methods to compare single-modal and multimodal four-class
decoding, confirming the advantage of multimodal fusion. We expect that the MIND dataset will facilitate the development of
data-driven methods for both standalone EEG-based and hybrid fNIRS--EEG
MI-BCI systems.

In existing public MI--BCI datasets, most paradigms use bilateral 2-class or 4-class MI tasks and are primarily based on single-modality EEG, such as BCI Competition IV~\cite{Brunner2008seta, lee2008B}, PhysioNet EEG-MI~\cite{Schalk2004BCI2000}, Cho 2017~\cite{Cho2017}, OpenBMI~\cite{Min2029Open}, and Kaya 2018~\cite{kaya2018large}, which are widely used benchmarks for algorithm development. With the shift toward practical applications, more realistic EEG datasets have been released, including cross-day settings (e.g., Cross-Session MI~\cite{ma2022large}), paradigms with interference tasks (e.g., MI under Distraction~\cite{brandl2020access}), and datasets designed for stroke rehabilitation (e.g., Acute Stroke MI~\cite{liu2024stroke} and Lower-limb Stroke Longitudinal~\cite{liu2025multi}). However, these datasets are still largely limited to bilateral paradigms or EEG-only recordings. Meanwhile, a small number of multimodal fNIRS--EEG and unilateral MI datasets have emerged. For example, Shin 2017~\cite{shin2017eegfnirs} provided a hybrid fNIRS--EEG dataset for left-/right-hand MI and mental arithmetic. OpenNeuro ds004022~\cite{seho2022} collected fNIRS--EEG signals during limb MI. Yi et al.~\cite{yi2025joints} released an fNIRS--EEG dataset for unilateral upper-limb multi-joint MI in Scientific Data. Rong et al.~\cite{Rong2024limbs} reported EEG decoding performance for unilateral upper-limb multi-class directional MI. In addition, some related unilateral multi-joint MI paradigms (e.g., Guo et al.~\cite{2024fusion}) have been studied but have not been released as complete public datasets. As summarized in Table~\ref{tab:dataset_comparison}, publicly available datasets that simultaneously satisfy the combination of unilateral upper limb, multi-class 2D directional commands, and multimodal fNIRS--EEG are still missing.

\begin{table*}[!ht]
\centering
\small
\setlength{\tabcolsep}{5pt}
\renewcommand{\arraystretch}{1.1}
\caption{Comparison of representative MI-BCI datasets and related studies. \cmark~indicates ``yes'' and \na~indicates ``no / not reported''.}
\label{tab:dataset_comparison}
\resizebox{0.8\linewidth}{!}{
\begin{tabularx}{\textwidth}{p{3cm}ccccX}
\toprule
\textbf{Dataset / Year}  & \textbf{fNIRS--EEG} & \textbf{Unilateral} & \textbf{Directional} & \textbf{Public} & \textbf{Notes} \\
\midrule
BCI-IV-2a (2008)                 & \na   & \na   & \na   & \cmark & L/R hand, foot, tongue \\
BBCI-IV-2b (2008)                & \na   & \na   & \na   & \cmark & L/R hand \\
PhysioNet EEG-MI (2004)          & \na   & \na   & \na   & \cmark & L/R; both fists / both feet \\
Cho (2017)                        & \na   & \na   & \na   & \cmark & L/R hand \\
OpenBMI (2019)                    & \na   & \na   & \na   & \cmark & MI / ERP / SSVEP \\
Kaya (2018)                       & \na   & \na   & \na   & \cmark & Up to 6 imagery tasks; includes 4-class MI settings \\
Cross-Session MI (2022)           & \na   & \na   & \na   & \cmark & L/R hand \\
MI under Distraction (2020)       & \na   & \na   & \na   & \cmark & Reach/MI with an interference task \\
Acute Stroke MI (2024)            & \na   & \cmark & \na   & \cmark & L/R hand-grip imagery (stroke) \\
Lower-limb Stroke Longitudinal (2025)
                                 & \na   & \cmark & \na   & \cmark & Lower-limb gait MI (with sensory cueing/stimulation) \\
TU Berlin hBCI (2017)             & \cmark& \na   & \na   & \cmark & L/R + MA (mental arithmetic) \\
OpenNeuro (2022)                  & \cmark& \na   & \na   & \cmark & Grasp $\rightarrow$ Twist $\rightarrow$ Reach $\rightarrow$ Lift \\
Guo (2024)                         & \cmark& \na   & \na   & \na    & L/R \\
Rong (2024)                        & \na   & \cmark & \cmark & \na    & Unilateral upper-limb 4-class MI (direction) \\
Yi (2025)                          & \cmark& \cmark & \na   & \cmark & Unilateral upper-limb multi-joint MI \\
Yang (2025)                        & \na   & \na   & \na   & \cmark & MI (L/R/feet) \\
\textbf{Proposed }                  & \cmark& \cmark & \cmark & \cmark & Unilateral upper-limb 4-class MI (direction) \\
\bottomrule
\end{tabularx}
}
\end{table*}

\section*{Methods}
\subsection*{Participants}
In this study, we recruited 30 healthy, right-handed participants
(18 males, 12 females; 19--25 years old) from college students. All participants had
normal or corrected-to-normal vision and reported no history of
neurological, psychiatric, or musculoskeletal disorders that could affect
the experimental results.
The study protocol was approved by the Ethics Committee. All procedures
followed institutional review board (IRB) guidelines and relevant privacy
regulations. Before the experiment, participants received a detailed
explanation of the study purpose, procedures, and requirements, including
being in good physical condition during the 24~hours before the experiment,
getting sufficient sleep, avoiding alcohol and caffeine, and washing their
hair in advance to ensure good signal quality. After confirming that they
understood the instructions, all participants voluntarily signed written informed consent forms.

Strict environmental controls were maintained throughout the EEG and fNIRS
data acquisition. We avoided strong magnetic fields in the experimental
area, minimized external sensory interference such as ambient noise and
light, and kept the room quiet and dim to reduce distractions. Additional
measures were taken to maximise participant comfort and to reduce
movement-related artefacts during recording. To minimise age-related variability and ensure consistent understanding of
the tasks, all participants were university students. The recruitment
procedure emphasised privacy and confidentiality, and clearly stated that
participation was voluntary. 

\subsection*{Experimental Paradigm}
\begin{figure}[ht]
	\centering
    \includegraphics[scale=0.5]{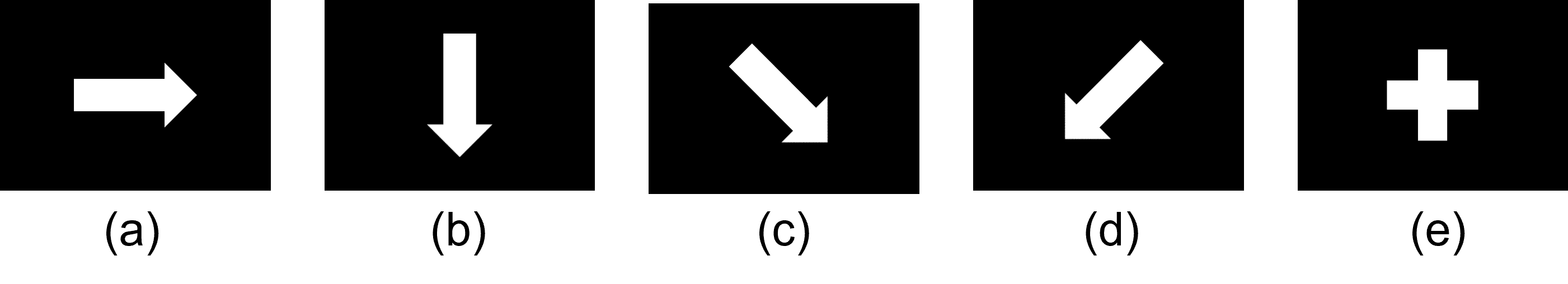}
    \caption{Representations of visual instruction to each unilateral limb task. (a) horizontal movement from left to right. (b) vertical movement from top to bottom. (c) diagonal movement from upper left to lower right. (d) diagonal movement from upper right to lower left (e) rest.}
	\label{Paragigm1}
\end{figure}

The experimental paradigm was designed to elicit cortical activation
patterns corresponding to different unilateral limb motor imagery (MI)
tasks. Specifically, four directional movement types were included:
horizontal left-to-right, vertical up-to-down, diagonal upper-left to
lower-right, and diagonal upper-right to lower-left. At the beginning of
each trial, participants received simultaneous visual and auditory cues.
For MI trials, the visual cue was presented as a 2~seconds white instructional
animation on a black background. For rest, a white fixation cross was
displayed on a black background together with text indicating the duration
of the upcoming rest period (Fig.~\ref{Paragigm1}).

\begin{figure}[ht]
	\centering
    \includegraphics[scale=0.5]{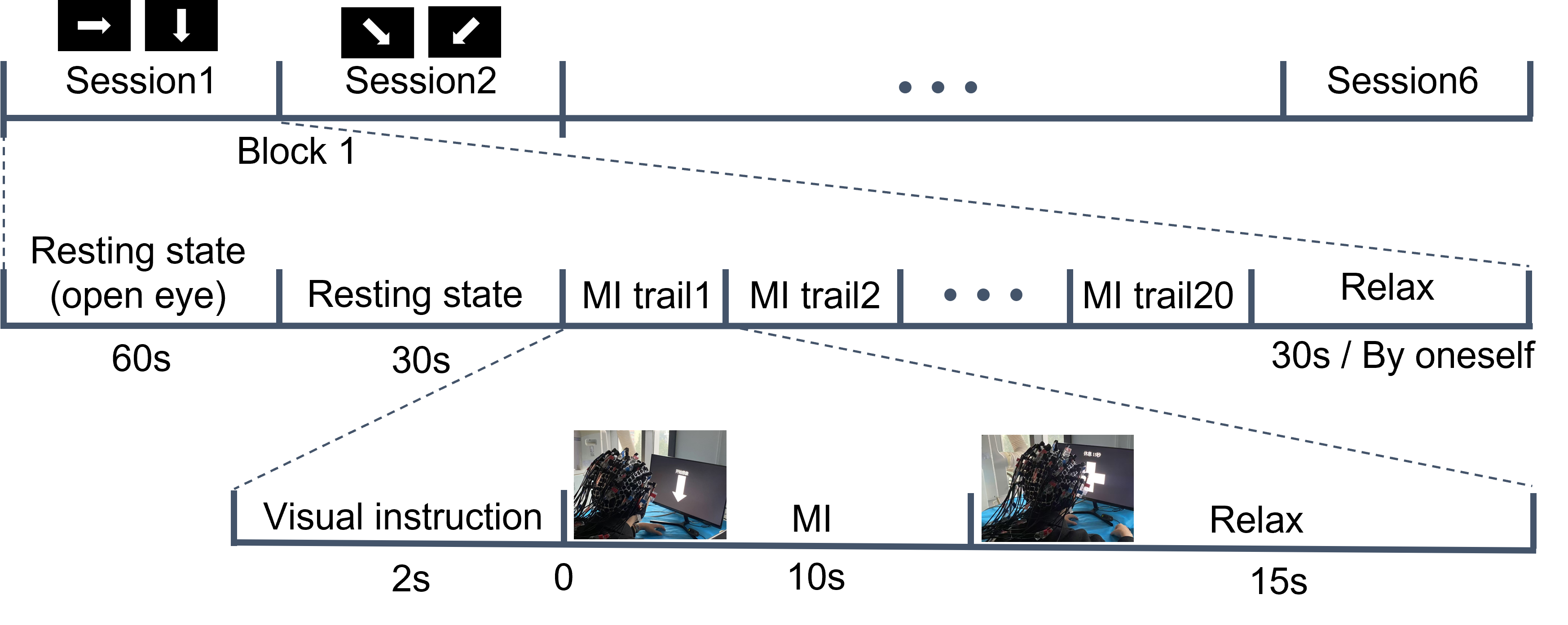}
    \caption{Experimental Paradigm of unilateral limb MI. Each subject completed three modules, with two sessions in each module, namely horizontal, vertical and two types of diagonal movements, totaling six sessions. Each session included a resting state and 20 MI sessions, making a total of 120 task sessions.}
	\label{Paragigm2}
\end{figure}
As shown in Fig.~\ref{Paragigm2}, each participant completed three blocks, and each block
consisted of two sessions, resulting in six sessions per participant. To
reduce potential bias caused by task-specific variability, horizontal and
vertical MI tasks were assigned to one session type, whereas the two
diagonal MI tasks were assigned to another session type. This grouping
strategy was intended to improve the stability and reproducibility of
neural responses across sessions.

Each block lasted ~11 min and included a 60~seconds eyes-open rest, a 60~seconds eyes-closed rest, and 20 MI trials. As illustrated in Fig.~\ref{Paragigm2}, each trial lasted
27~seconds. The trial began with a 2~seconds visual cue, at 0~second, visual and auditory
cues were presented synchronously, with a white arrow on a black background
indicating the required imagery direction and the Chinese characters
“开始想象（Start imagining）”, accompanied by a short “ding” sound. This was followed by a 10~seconds MI execution period, during which participants were
instructed to continuously imagine performing the cued limb movement and
were encouraged to mentally repeat the movement 2--4 times within this
interval. After the MI phase, a white fixation cross on a black background
was shown, indicating the start of a 15~seconds inter-trial rest period. Each
session contained 20 MI trials, giving a total of 120 trials per
participant across the six sessions. The four MI task types (horizontal,
vertical and the two diagonal directions) were each performed 30 times. To
ensure data quality and minimise confounding factors, extraneous visual or
auditory stimuli were strictly avoided throughout the experiment, providing
a controlled and consistent sensory environment.

\subsection*{Data collection and Preprocessing}
\subsubsection*{Data collection}
\begin{figure}[ht]
	\centering
    \includegraphics[scale=0.35]{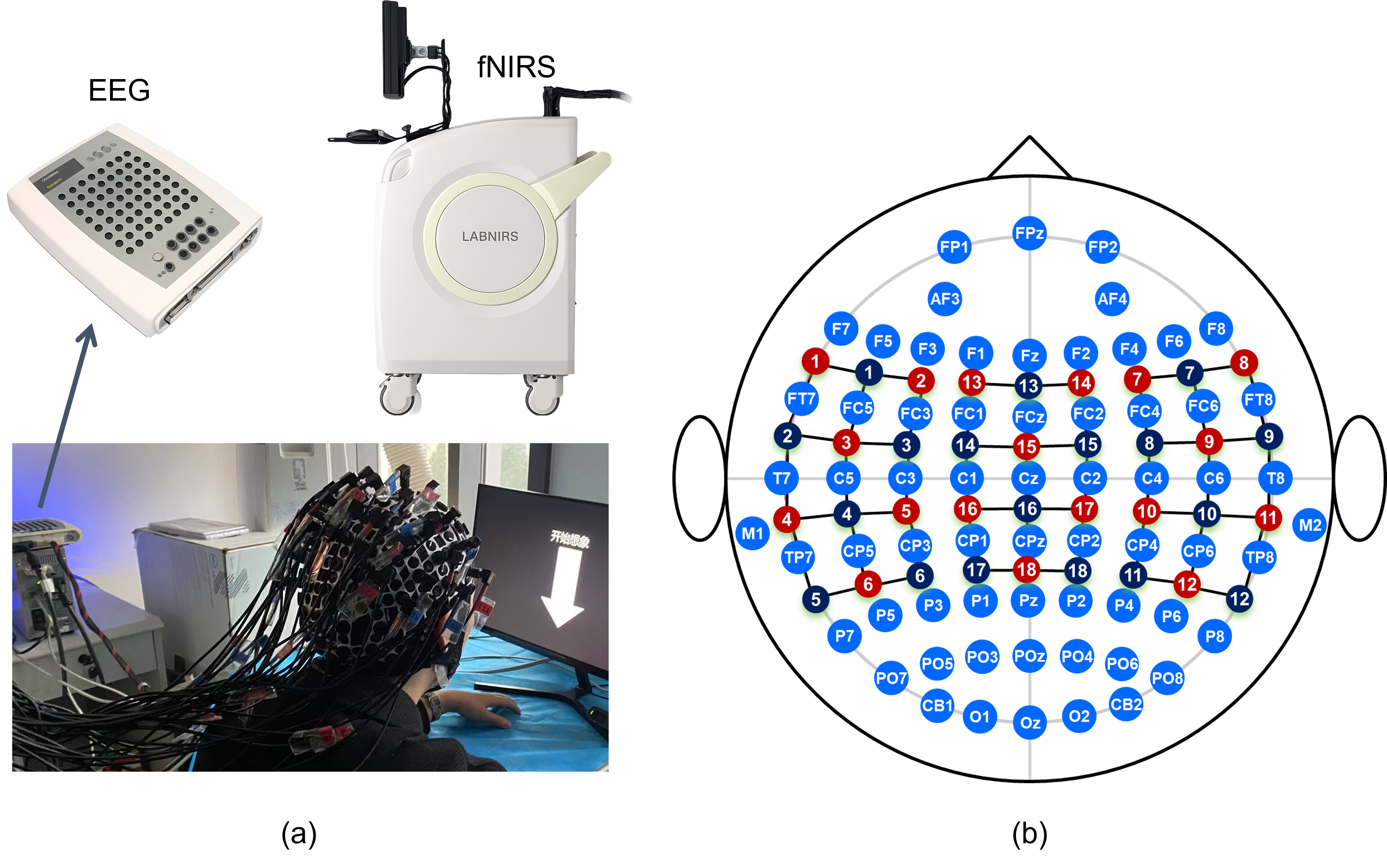}
    \caption{The instrumentation used in eeg data collection. (a) The EEG cap and signal  amplifier. (b) The electrode positions on the EEG cap.}
	\label{Instrumentation}
\end{figure}

An integrated fNIRS--EEG cap and signal synchroniser developed by
RSx\&r was used to enable joint acquisition and precise temporal
alignment of both modalities (Fig.~\ref{Instrumentation}(a). The channel layout for the
integrated fNIRS--EEG recording is shown in Fig.~\ref{Instrumentation}(b). All recordings
were conducted in a controlled environment to minimise ambient light
interference and movement-related artefacts, and participants were
instructed to remain as still as possible throughout the experiment. The experimental paradigm was designed and implemented using E-Prime~3.0
and its development environment E-Studio~\cite{E-Prime}. Timing synchronisation between
the fNIRS--EEG recording system and the experimental paradigm was achieved
via digital triggers sent from E-Prime to the amplifiers, ensuring precise
alignment between task events and physiological data. 

EEG signals were recorded using a solid electrode cap equipped with
64 Ag/AgCl electrodes arranged according to the international 10--20
system (Neuroscan). This type of electrode cap provides high current
density, strong anti-interference capability and low contact impedance.
Signal acquisition was performed with a Neuroscan synAmps2 wireless
amplifier, which supports real-time impedance monitoring and operates
in a wireless transceiver mode (Fig.~\ref{Instrumentation}(a)). The reference electrode (REF)
was placed over the parietal region, and the ground electrode (GND)
was positioned on the forehead. EEG signals were sampled at 1000~Hz.
During data collection, electrode impedance was maintained below
10~k$\Omega$ to ensure signal quality. The spatial distribution of
electrodes on the EEG cap is illustrated in Fig.~\ref{Instrumentation}(b).

fNIRS signals were recorded using a
continuous-wave LABNIRS system (Shimadzu Corp., Kyoto, Japan) to measure
task-related hemodynamic responses. The device operated at three
wavelengths (780, 805 and 830~nm). In total, 18 emitters and 18 detectors
were used, forming 51 measurement channels; the spatial layout of these
channels is shown in Fig.~\ref{Instrumentation}(b). Before each recording, the optical pathways
were adjusted and checked to ensure that the signal quality of every
channel met the manufacturer's recommended criteria. The LABNIRS system
outputs concentration changes in oxygenated, deoxygenated and total
hemoglobin, and all fNIRS preprocessing was performed on these
concentration time series.

In the present study, fNIRS signals were thus acquired with a 51-channel
continuous-wave LABNIRS system (Fig.~\ref{Instrumentation}(a)). The optode arrangement followed
the international 10--20 system, with source--detector pairs placed mainly
over the prefrontal, parietal, temporal and sensorimotor cortices to
capture movement-related hemodynamic activity. The inter-optode distance
was set to 30~mm to ensure appropriate cortical penetration depth, and all
optodes were secured using an elastic cap to maintain stable contact
throughout the sessions. Data were sampled at 47.62~Hz and recorded
continuously during the entire experiment.

\subsubsection*{Data preprocessing}
\textbf{EEG preprocessing:} EEG data were preprocessed using the
MNE-Python toolbox~\cite{mne} to reduce noise and improve signal quality for
subsequent analyses. First, a 0.5--50~Hz band-pass filter was applied to
remove slow drifts and high-frequency noise, followed by a 50~Hz notch
filter to suppress power-line interference. Channels with poor signal
quality were then reconstructed using spherical spline interpolation. Next, data segments corresponding to the four MI classes and the resting
state were extracted: each MI segment lasted 25~seconds (5~seconds pre-cue interval,
10~seconds MI execution period, and 10~seconds post-imagery rest), and the resting
segment lasted 60~seconds. A baseline window from $-2$~second to 0~second before cue onset was used for baseline correction to remove DC offsets. The sampling rate
was subsequently reduced from 1000~Hz to 250~Hz to decrease computational
load while preserving neurophysiologically relevant frequency components.
To further denoise the data, we used the EEGLAB tool in MATLAB for independent component analysis (ICA) to decompose the EEG into 30 independent components. ICLabel automatic classification tool was applied with probability-based thresholds
to identify and remove artefactual components (muscle activity, eye
movements and channel noise). Finally, the preprocessed EEG data were
exported in MATLAB-compatible $.mat$ format for downstream decoding
and statistical modelling.

\textbf{fNIRS preprocessing:} fNIRS data were preprocessed using the
MNE-Python toolbox~\cite{mne} to reduce noise and extract reliable hemoglobin concentration changes. First, a third-order Butterworth band-pass filter (0.02--0.2~Hz) was applied to the raw signals to remove slow drifts and
high-frequency noise, while preserving task-related hemodynamic fluctuations. Using
event markers, task-related time windows (e.g.\ from $-5$~second to 25~second around
stimulus onset) were then extracted and segmented into single trials,
yielding a three-dimensional data array with dimensions
trial~$\times$~channel~$\times$~time. A baseline interval from $-5$~second to
0~second before task onset was used for baseline correction to remove slow
offsets and drift. Finally, the fNIRS data were down-sampled to 10~Hz to
reduce computational load while retaining physiologically meaningful
information. The preprocessed HbO and HbR time series were exported in $.csv$
format for subsequent analysis, providing a complete fNIRS preprocessing
pipeline that ensures signal quality suitable for downstream analyses and
decoding experiments.

\section*{Data Records}

\begin{figure}[!h]
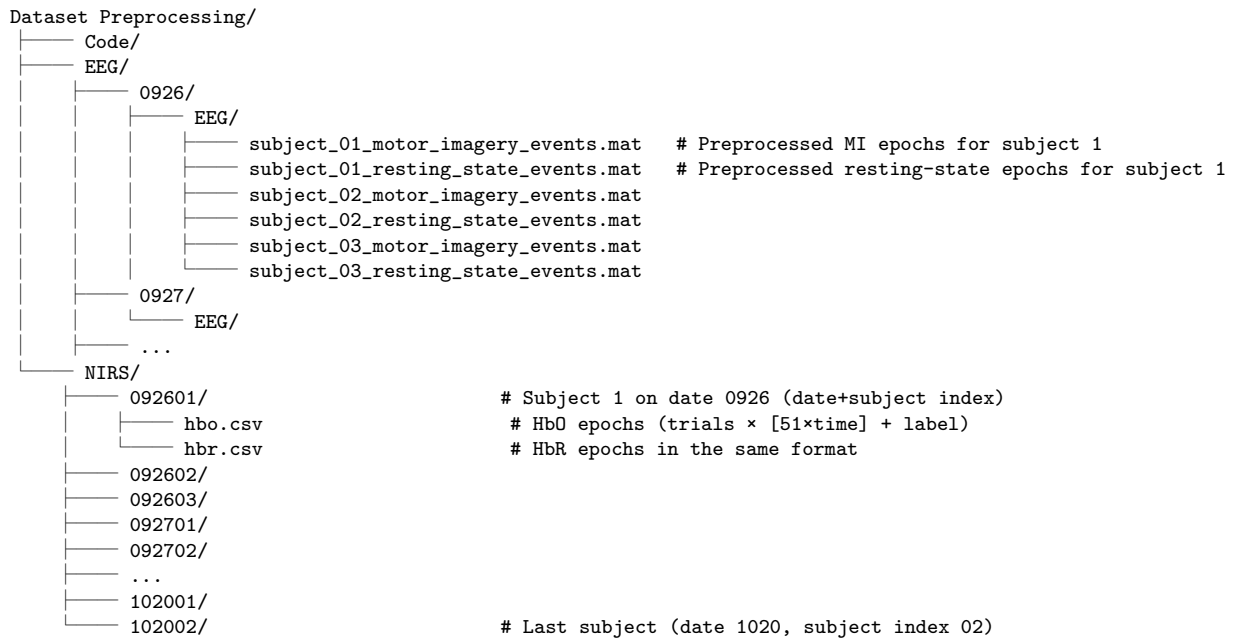

  \centering
  \scriptsize
  \begin{Verbatim}[breaklines=true, breakanywhere=true]
Dataset Preprocessing/
├── Code/
├── EEG/
│   ├── 0926/
│   │   ├── EEG/
│   │   │   ├── subject_01_motor_imagery_events.mat   # Preprocessed MI epochs for subject 1
│   │   │   ├── subject_01_resting_state_events.mat   # Preprocessed resting-state epochs for subject 1
│   │   │   ├── subject_02_motor_imagery_events.mat
│   │   │   ├── subject_02_resting_state_events.mat
│   │   │   ├── subject_03_motor_imagery_events.mat
│   │   │   └── subject_03_resting_state_events.mat
│   ├── 0927/
│   │   └── EEG/
│   ├── ...
└── NIRS/
    ├── 092601/                          # Subject 1 on date 0926 (date+subject index)
    │   ├── hbo.csv                      # HbO epochs (trials × [51×time] + label)
    │   └── hbr.csv                      # HbR epochs in the same format
    ├── 092602/
    ├── 092603/
    ├── 092701/
    ├── 092702/
    ├── ...
    ├── 102001/
    └── 102002/                          # Last subject (date 1020, subject index 02)
  \end{Verbatim}
  \caption{Directory structure of the Preprocessed multimodal EEG–fNIRS dataset.}
  \label{tree3}
\end{figure}

The recordings have been released on a public data repository (ScienceDB) and are openly accessible. As shown in Fig.~\ref{tree2} and Fig.~\ref{tree3}, the MIND dataset in the repository consists of two parts: 
(1) the raw data are stored in the folder $Dataset raw/$; 
(2) the processed data are stored in the folder $Dataset Preprocessing/$. 
In addition, a file named $dataset\_description.md$ is provided, which summarizes all key information for reuse, including the sampling rates, the meaning of each variable/field, the task sequence, data loading instructions, and other relevant details. The entire MIND dataset is publicly available on the ScienceDB platform. 
The released data are licensed under the Creative Commons Attribution 4.0 International License (CC BY 4.0).

\begin{figure}[ht]
  \centering
  \scriptsize
  \begin{Verbatim}[breaklines=true, breakanywhere=true]
Dataset_raw/   # Raw data recording folder
├─ 0926/                        # Acquisition date
│   ├─ EEG/                     # Raw EEG data folder
│   │   ├─ Acquisition 05.dat   # Continuous EEG data (including metadata and events)
│   │   ├─ Acquisition 05.ceo   # Project-level metadata: experiment configuration, \\number of channels, etc.
│   │   ├─ Acquisition 05.dap   # Acquisition parameters: sampling rate, gain, etc.
│   │   ├─ Acquisition 05.rs3   # Spatial and device configuration: electrode/sensor locations
│   │   ├─ Acquisition 06.dat
│   │   ├─ Acquisition 06.ceo
│   │   ├─ Acquisition 06.dap
│   │   ├─ Acquisition 06.rs3
│   │   └─ ...                  # Each .dat file contains one block group for one subject; each subject has three consecutive .dat files
│   ├─ fNIRS/                   # Raw fNIRS data folder
│   │   ├─ 0926-001.csv         # Onset times and task labels for this block
│   │   ├─ 0926-001.txt         # Continuous fNIRS data (each subject has three consecutive .txt files)
│   │   ├─ 0926-002.csv
│   │   ├─ 0926-002.txt
│   │   ├─ 0926-003.csv
│   │   ├─ 0926-003.txt
│   │   ├─ 0926-101.csv
│   │   ├─ 0926-101.txt
│   │   └─ ...                  # Each subject has three consecutive .txt and .csv files: xx-001, xx-002, xx-003
│   └─ ...
├─ 0927/                        # Other acquisition dates
│   ├─ EEG/
│   ├─ fNIRS/
│   └─ ...
├─ [Other dates]/               # Additional acquisition dates
│   └─ ...
├─ NIRS Event/                  # Manually curated event labels for fNIRS
│   ├─ 0927/
│   │   ├─ Subject_1_events.csv
│   │   ├─ Subject_2_events.csv
│   │   └─ Subject_3_events.csv
│   ├─ 0928/
│   │   └─ ...
│   ├─ 1004/
│   └─ 1020/
└─ Optical.txt                  # Source–detector and channel coordinate information for fNIRS optodes
  \end{Verbatim}
  \caption{Directory structure of the raw multimodal fNIRS--EEG dataset.}
  \label{tree2}
\end{figure}
This dataset contains simultaneously recorded  EEG 
and fNIRS signals. The structure of the raw data is illustrated in Fig.~\ref{tree2}. Data are organised by acquisition date, and for each date there are two subfolders: $EEG/$ and $fNIRS/$.
In the $EEG/$ folder for a given date, four file types are present:
$.dat$, $.ceo$ $.dap$ and $.rs3$. The
continuous raw EEG signals are stored in the $.dat$ files. For each
participant, three consecutive $.dat$, $.ceo$, $.dap$
and $.rs3$ files are provided, ordered by increasing acquisition
number. Each $.dat$ file contains two consecutive sessions, so that
each participant has three consecutive $.dat$ files in total. For
example, in $0926/EEG$, the files $Acquisition 05.dat$,
$Acquisition 06.dat$ and $Acquisition 07.dat$ correspond to
the three block groups of the first participant recorded on 26~September.
The $.dat$ files can be loaded using the Neuroscan reader in
MNE-Python into a $Raw$ object (here denoted as $data$). The
$data.info$ structure provides channel locations, sampling
frequency, channel names and other metadata, and the event annotations
encode the task structure. In the event codes, ``1" denotes the
eyes-open resting state,  ``2" denotes the eyes-closed resting state,
and ``4"-- ``7" denote the four MI tasks: left-to-right,
up-to-down, upper-left to lower-right and upper-right to lower-left,
respectively.
 
In the $fNIRS/$ folder for each date, two file types are provided:
$.csv$ and $.txt$. Files are named using the pattern
$[date]-[ID]$, where $[ID]$ is a three-digit code combining
participant and block indices; within $[ID]$, the first digit
indicates the participant index for that date, and the last digit indicates
the block (session group) index. For example, $0926-001.csv$ and
$0926-001.txt$ correspond to the first block of the first
participant on 26~September; each participant has three such block IDs,
e.g.\ $0926-001$, $0926-002$ and $0926-003$. Each
$.txt$ file contains the raw fNIRS measurements (oxyHb, deoxyHb,
totalHb, Abs780nm, Abs805nm and Abs830nm) together with channel
information, while the corresponding $.csv$ file stores the event
onsets and labels. For the dates $0927$, $0928$, $1004$ and $1020$, fNIRS events were manually annotated. In these cases, the $Mark$
column in the $.txt$ data table denotes the time points at which the
experimenter pressed the marker key. A helper function $hand\_mrk()$
was used to read the $Time(sec)$ values where $Mark == 1$,
thereby obtaining the marker time series. Based on the experimental
paradigm (fixed block durations), each marker press was converted into an
accurate event time ($mrk\_time$) and assigned a task label. All
derived events were then saved to $NIRS Event/[date]/Subject\_X\_events.csv$,
representing the combined events for the three blocks of participant~X on
that date. In addition, for the first participant on $0928$, the
second block was split into two files ($0928-002\_1$ and
$0928-002\_2$) due to a temporary device interruption. We read both
$.txt$ segments separately and re-concatenated them after aligning
to the unified event table $Subject\_1\_events.csv$. For the date
$1004$, the second block of the first participant was corrupted, so
only the first and third blocks were retained.

Fig.~\ref{tree3} shows the structure of the preprocessed data. Preprocessed EEG
signals are stored in MATLAB $.mat$ files, whereas preprocessed
fNIRS HbO and HbR time series are stored as $.csv$ files. For each
participant, two preprocessed EEG files and two preprocessed fNIRS files
are provided. EEG files follow the naming convention
$[date]/EEG/subject\_0^*/motor\_imagery\_events.mat$ and
$[date]/EEG/subject\_0^*/resting\_state\_events.mat$, where
$0^*$ denotes the participant index for that date, and
$motor\_imagery\_events$ and $resting\_state\_events$
correspond to MI and resting-state data, respectively. fNIRS files are
named as $[data]0^*/hbr.csv$ and $[data]0^*/hbo.csv$, where $[data]0^*$
encodes the participant's date and index. The preprocessed EEG
$.mat$ files can be directly loaded in MATLAB and are compatible
with Python via $scipy.io.loadmat()$, while the fNIRS $.csv$
files can be read using standard functions such as $pandas.read\_csv()$.

\section*{Technical Validation}
\subsection*{EEG Data Analysis}
\textbf{Time–domain analysis (ERD/ERS).}
Event–related desynchronization/synchronization  (ERD / ERS) was used to
quantify changes in the sensorimotor rhythms that are closely related to
motor imagery, focusing on the $\alpha$ (8--12~Hz) and $\beta$ (13--30~Hz)
bands~\cite{erders}. The computation proceeded as follows: (1) the EEG signals were
band–pass filtered in the $\alpha$ and $\beta$ bands; (2) the filtered
signal was squared at each sample point to obtain instantaneous power;
(3) for each MI condition, power time series from all trials were summed
and averaged; (4) a sliding time window was applied to obtain a
smoothed power curve; (5) ERD/ERS was then expressed as
\begin{equation}
  \mathrm{ERD/ERS}(t) = \frac{A(t)-R}{R}\times 100\% ,
\end{equation}
where $A(t)$ denotes the task–related power at time $t$ and $R$ denotes
the baseline power.

\textbf{Frequency–domain analysis (ERSP).}
Event–related spectral perturbation (ERSP) was used to characterise
changes in spectral energy during MI~\cite{ERSP}. ERSP is an EEG analysis method
that examines how oscillatory power at different frequencies changes in
response to specific events or tasks, providing a time–frequency
representation relative to a baseline period, which is typically a rest
or quiet state. In this study, ERSP was computed as follows: (1)
preprocessed EEG data were decomposed in the time–frequency domain using
the short–time Fourier transform (STFT); (2) spectral power was computed
for each time–frequency point; (3) the interval preceding MI onset was
used as the baseline; (4) ERSP was obtained as the average spectral
energy across trials:
\begin{equation}
  \mathrm{ERSP}(f,t) = \frac{1}{n}\sum_{k=1}^{n} \left| F_k(f,t) \right|^{2},
\end{equation}
where $F_k(f,t)$ is the complex STFT value for trial $k$ at frequency
$f$ and time $t$, and $n$ is the number of trials.

\begin{figure}[ht]
  \centering
  \resizebox{0.75\textwidth}{!}{ 
    \begin{minipage}{\textwidth}
      \centering
      \subfloat[Left to right ERD/ERS]{
        \includegraphics[width=0.45\textwidth]{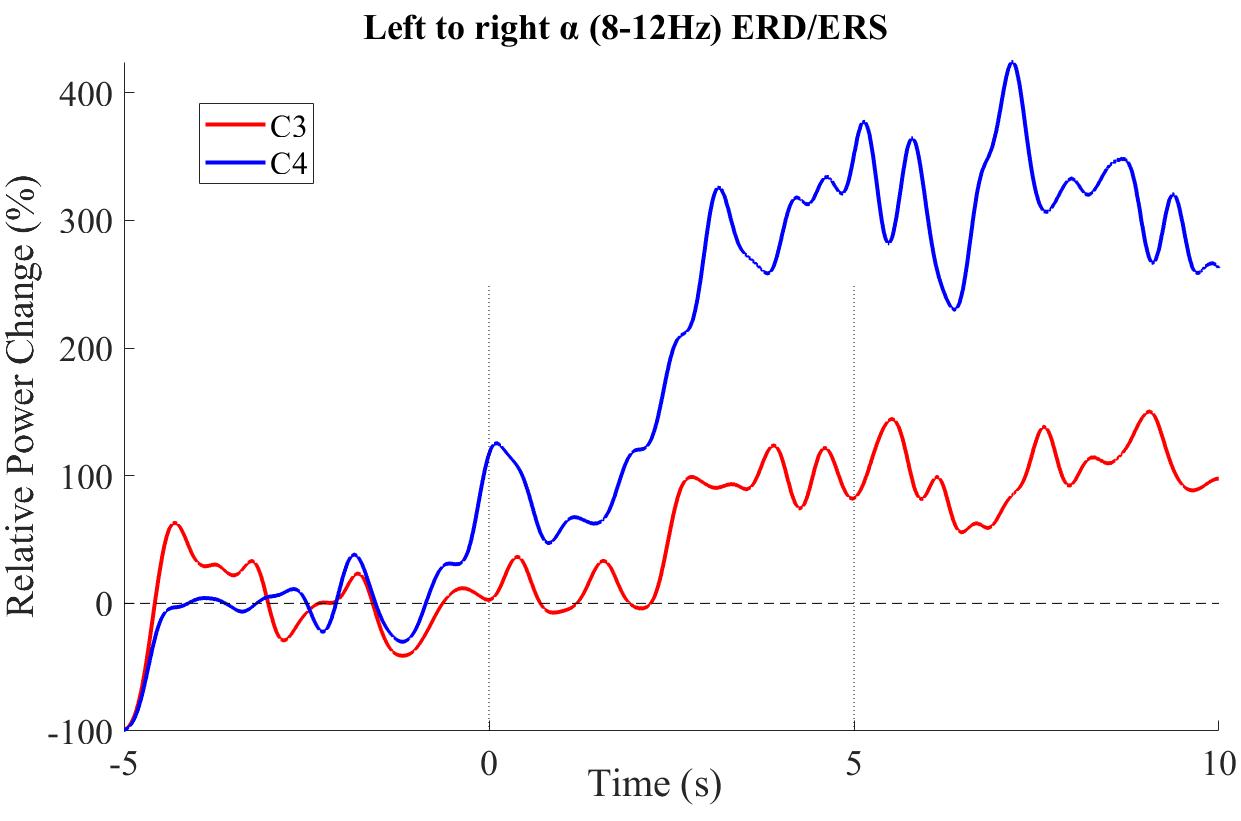}%
      }\hfill
      \subfloat[Top to bottom ERD/ERS]{
        \includegraphics[width=0.48\textwidth]{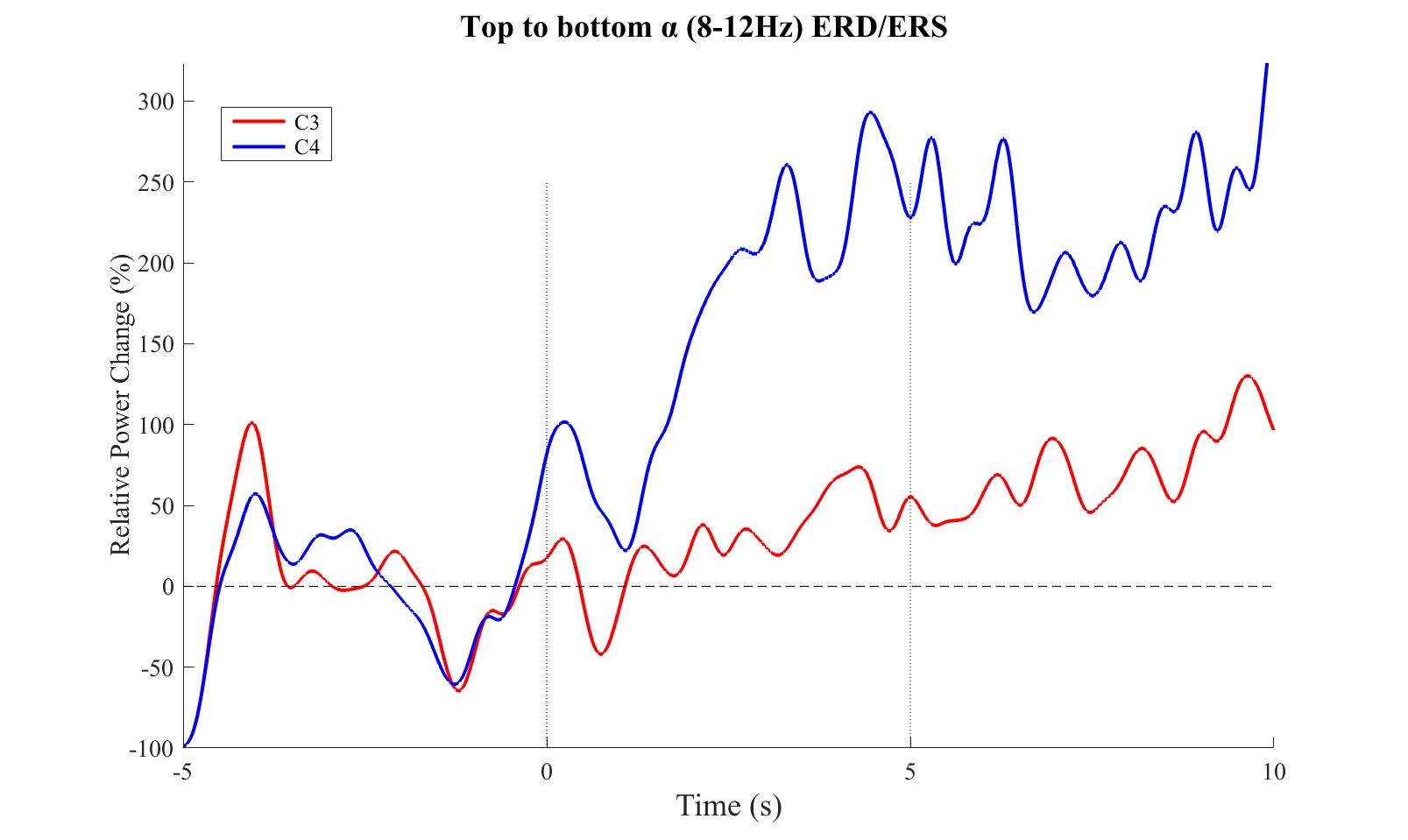}%
      }\\[2mm]
      \subfloat[Upper left to lower right ERD/ERS]{
        \includegraphics[width=0.47\textwidth]{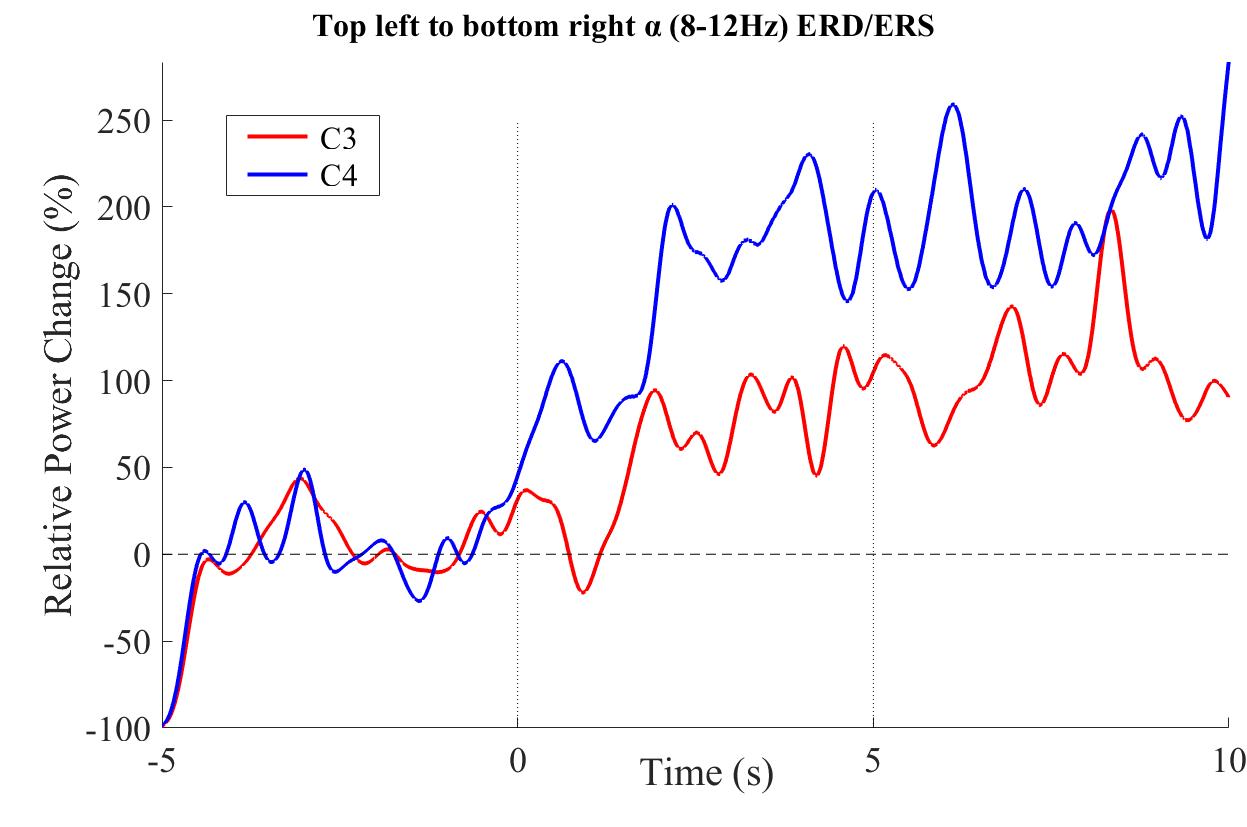}%
      }\hfill
      \subfloat[Upper right to lower left ERD/ERS]{
        \includegraphics[width=0.47\textwidth]{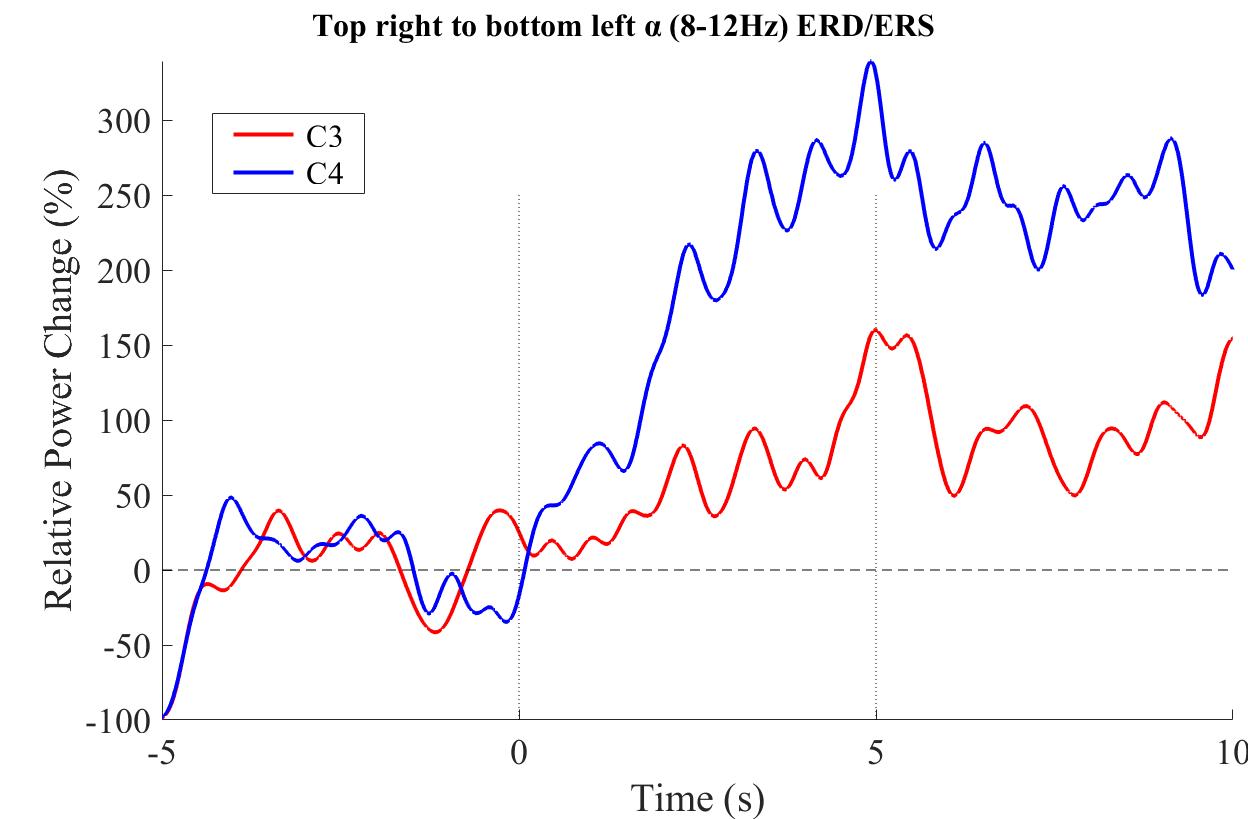}%
      }
    \end{minipage}
  }
  \caption{Four MI tasks: ERD/ERS curves of C3 and C4 in the
  $\alpha$ band (8--12 Hz).}
  \label{erders_alpha}
\end{figure}

\begin{figure}[!h]
  \centering
  \resizebox{1.0\textwidth}{!}{ 
    \begin{minipage}{\textwidth}
      \centering
      \subfloat[ERSP at C3 and C4 (Left-to-Right MI)]{
        \includegraphics[width=0.5\textwidth]{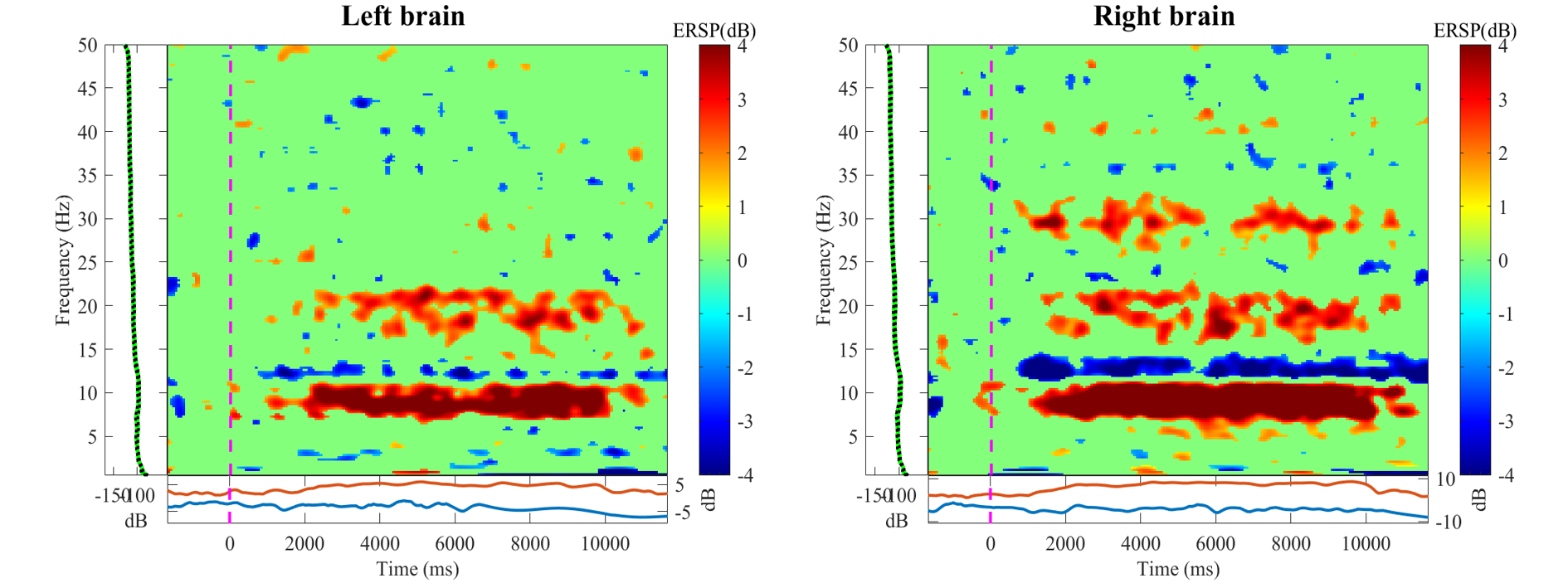}%
      }
      \subfloat[ERSP at C3 and C4 (Top-to-Bottom MI)]{
        \includegraphics[width=0.5\textwidth]{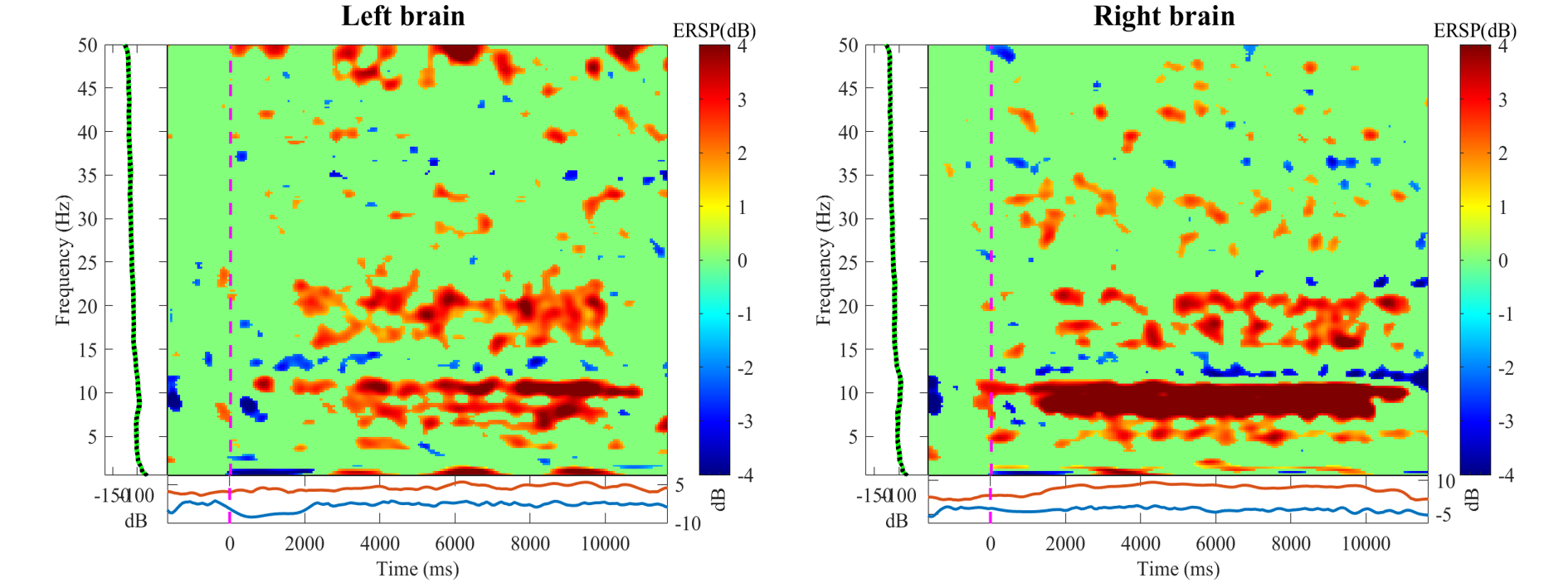}%
      }\\[2mm]
       \subfloat[ERSP at C3 and C4 (Upper Left-to-lower Right MI)]{
        \includegraphics[width=0.5\textwidth]{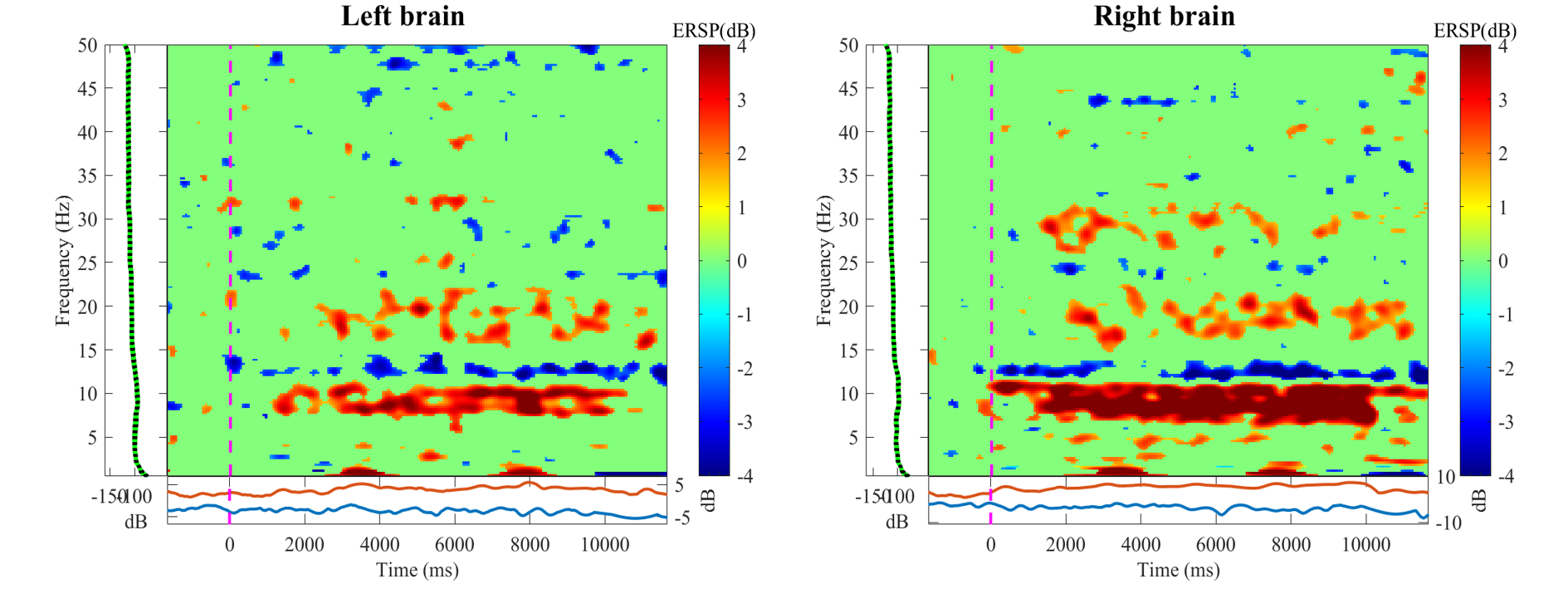}%
      }
      \subfloat[ERSP at C3 and C4 (Upper right-to-lower Left MI)]{
        \includegraphics[width=0.5\textwidth]{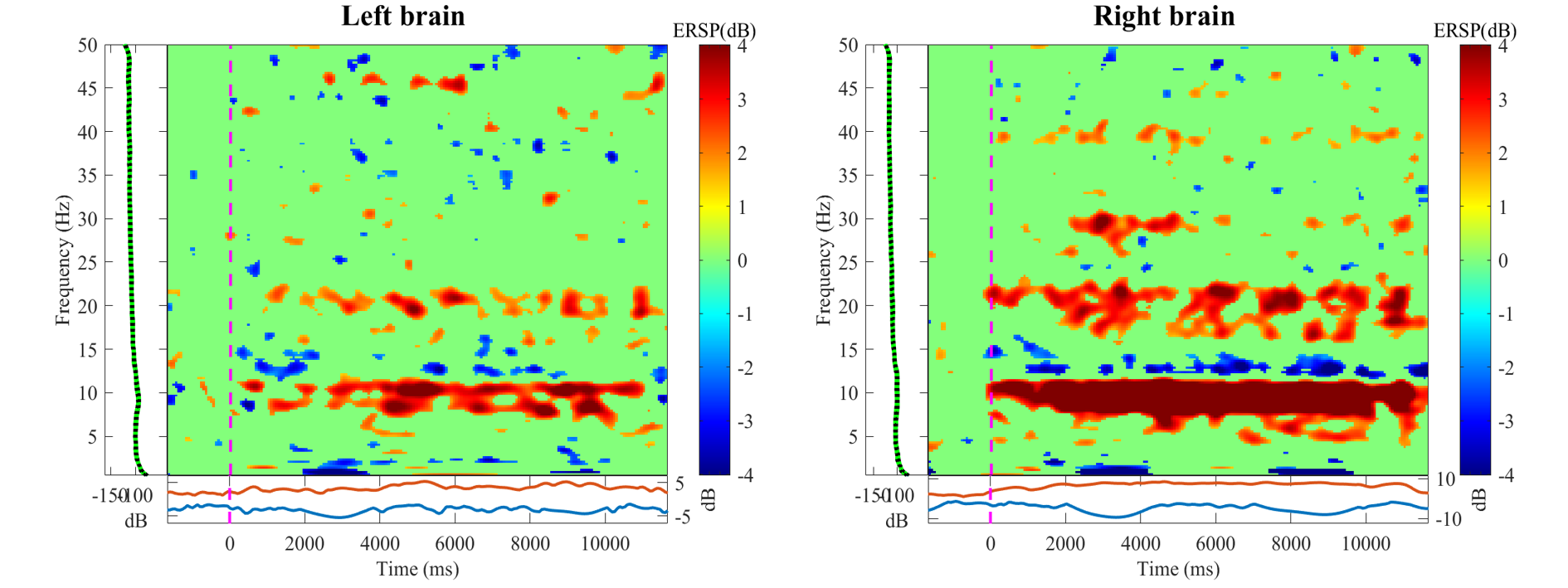}%
      }
    \end{minipage}
  }
  \caption{Average ERSP maps (8--30 Hz) over left and right sensorimotor areas(C3 and C4) during four MI tasks.}
  \label{ersp}
\end{figure}

\begin{figure}[ht]
	\centering
    \includegraphics[scale=1.0]{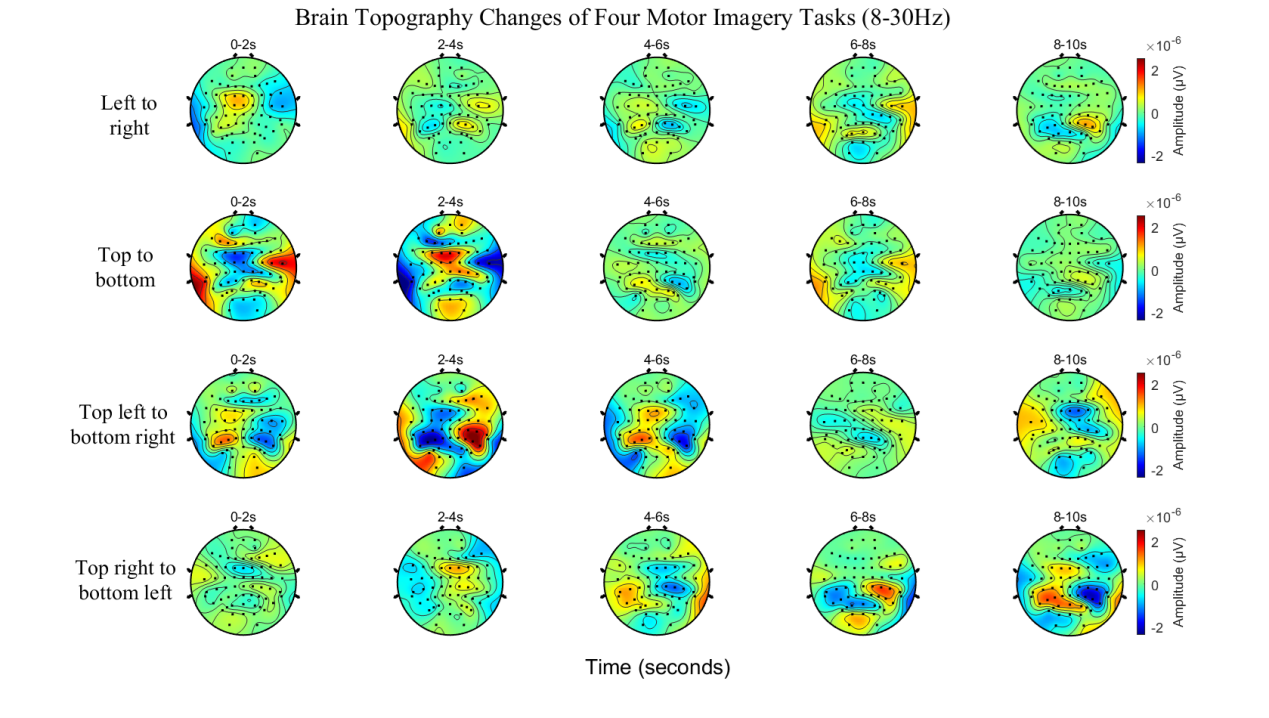}
    \caption{Topographic distribution of EEG for four directional motor imagery tasks.}
	\label{Topoeeg}
\end{figure}

Fig.~\ref{erders_alpha} shows the ERD/ERS curves of the $\alpha$ rhythm at channels C3 and C4 for the four MI tasks in Subject~1. Around the task onset (0~s), a clear power change appears at C4, and during the 0--10~s MI period the power at C4 remains higher than that at C3. This pattern is consistent with the typical contralateral--ipsilateral effect in actual movement, where activation of the contralateral sensorimotor cortex is accompanied by $\alpha$-power suppression, while the ipsilateral side shows a relative power increase.

Fig.~\ref{ersp} shows the ERSP in the 0.5--50~Hz frequency band for four MI tasks: (a) Left-to-Right MI, (b) Top-to-Bottom MI, (c) Upper Left-to-Lower Right MI, and (d) Upper Right-to-Lower Left MI. The figure demonstrates that after the event onset in all four tasks, the C3 channel shows significant power enhancement (red region) in the 8–12 Hz and 15–20 Hz frequency bands, while power suppression (blue region) occurs in the 12–15 Hz rhythm. Additionally, the C4 channel on the right side shows significant power enhancement (red region) in the 8–12 Hz, 15–20 Hz, and 30 Hz frequency bands, which is stronger than the left C3 channel, indicating contralateral dominance.

Fig.~\ref{Topoeeg} shows the time-varying scalp topographies in the 8--30~Hz band for the four motor imagery (MI) tasks. From top to bottom, each row
corresponds to one task condition: left-to-right, top-to-bottom,
upper-left-to-lower-right, and upper-right-to-lower-left, respectively.
Each column represents a different time window from 0--2~s to 8--10~s
after cue onset. Colors indicate power changes (in $\mu$V$^{2}$), with
red denoting power increases and blue denoting power decreases. The
maps reveal that different MI directions evoke distinct activation
patterns in specific regions (especially around the central
sensorimotor cortex) and show clear temporal evolution, reflecting the
spatial–temporal response of cortical areas to different movement
directions. A contralateral lateralization pattern can be observed in
the spatial distribution, whereas within-hemisphere differences between
the four tasks are subtle and cannot be reliably distinguished by
visual inspection alone.

\subsection*{fNIRS Data Analysis}

\begin{figure}[!h]
  \centering
  \resizebox{0.85\textwidth}{!}{ 
    \begin{minipage}{\textwidth}
      \centering
      \subfloat[Left to right]{
        \includegraphics[width=0.45\textwidth]{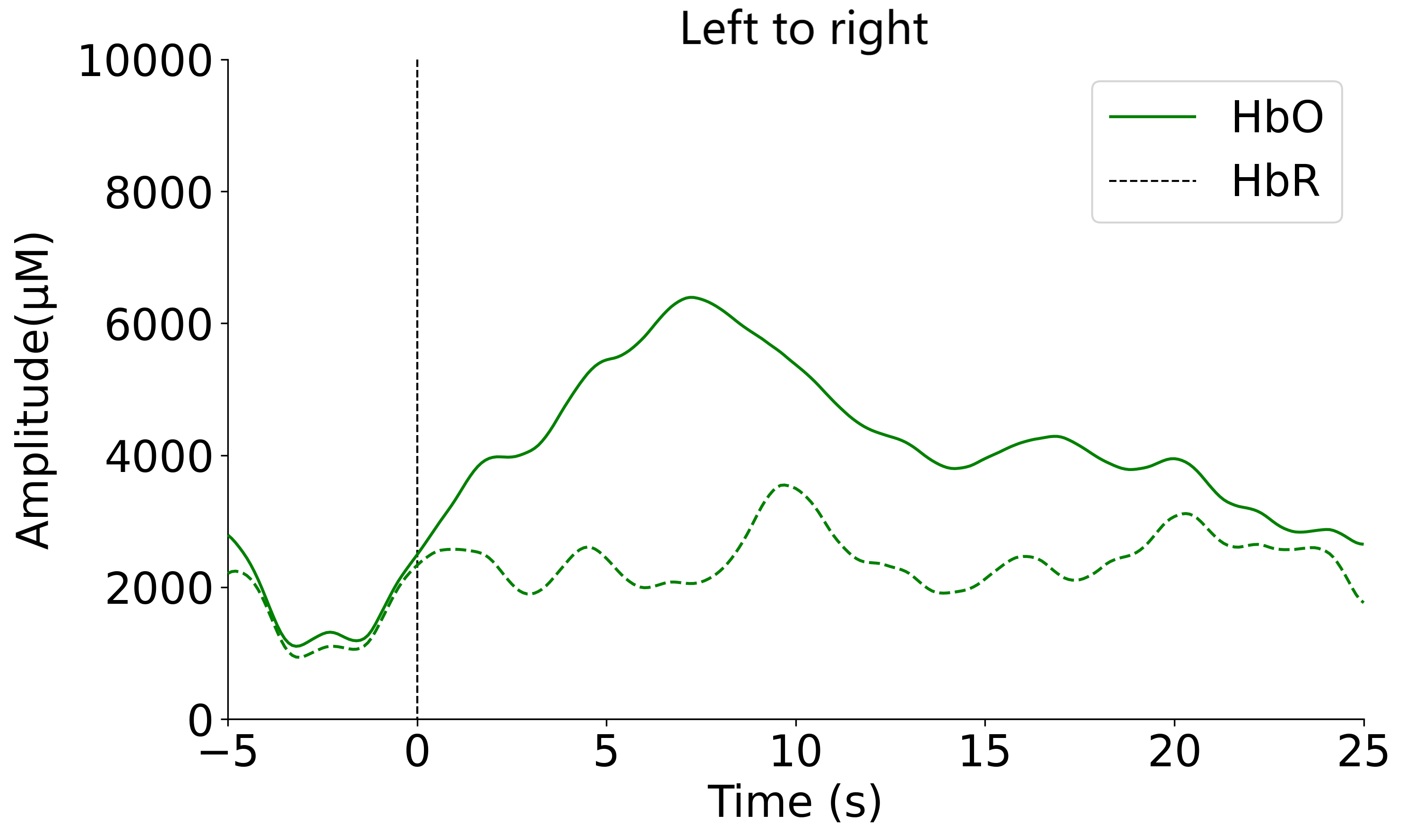}%
      }\hfill
      \subfloat[Top to bottom]{
        \includegraphics[width=0.45\textwidth]{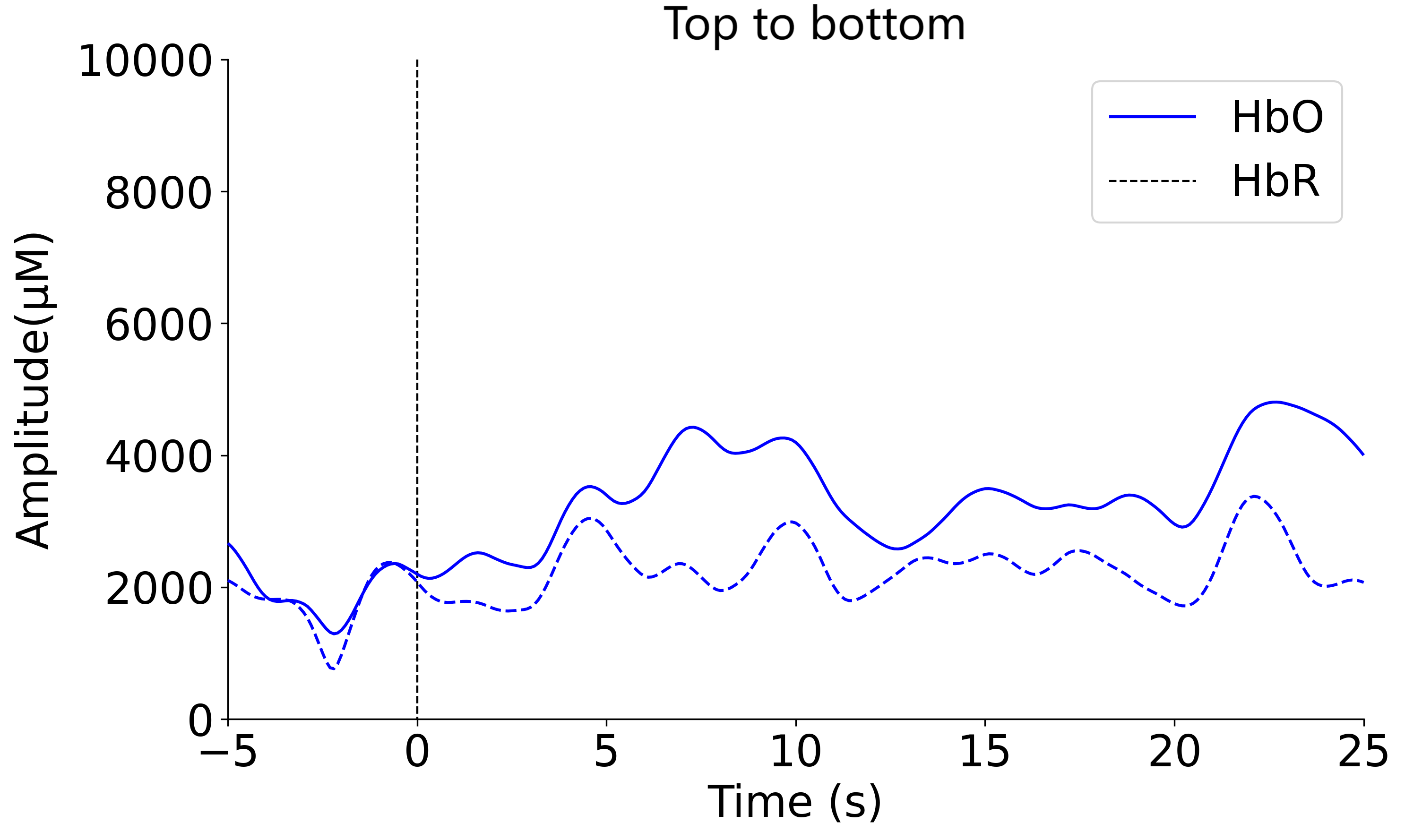}%
      }\hfill
      \subfloat[Upper left to lower right]{
        \includegraphics[width=0.45\textwidth]{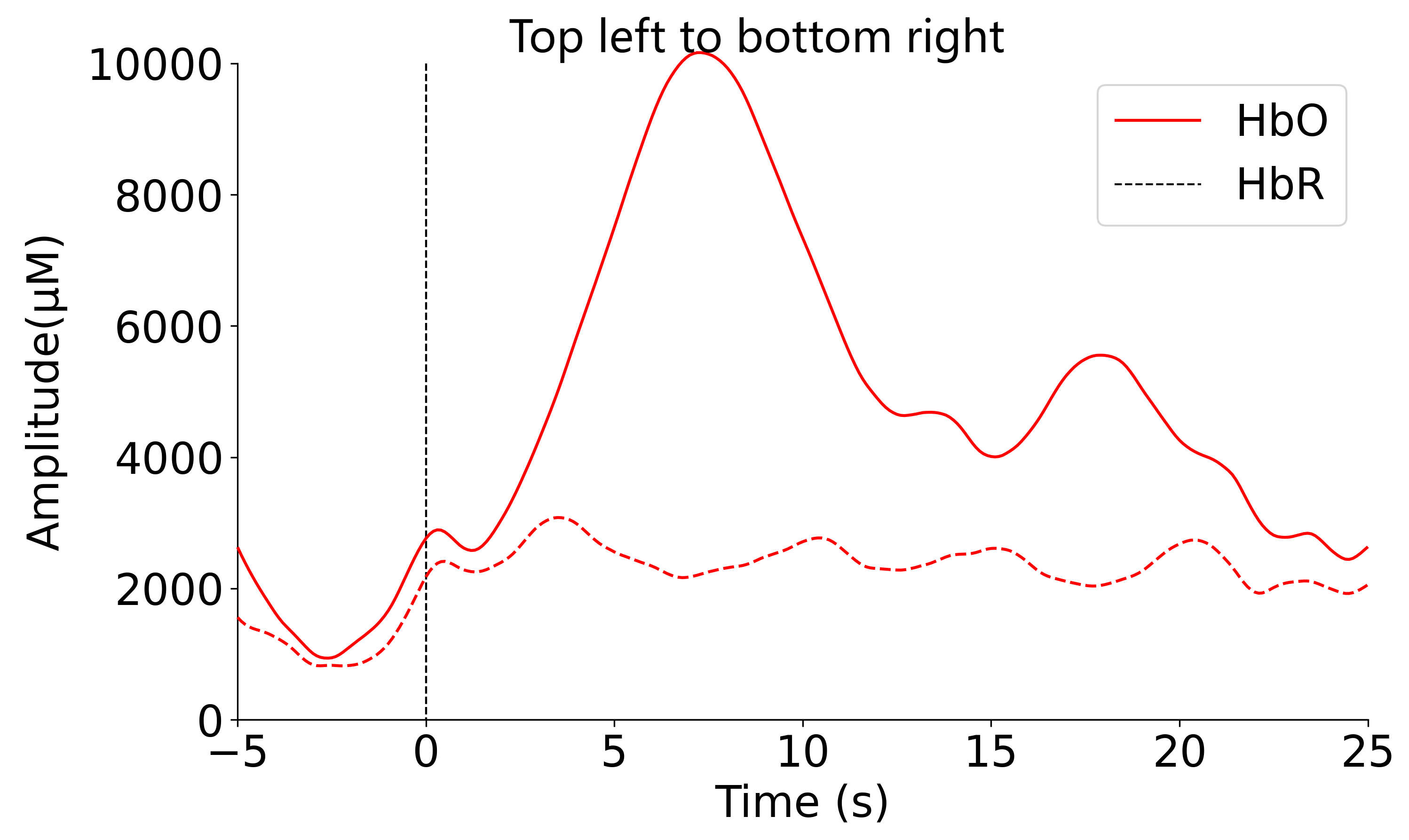}%
      }\hfill
      \subfloat[Upper right to lower left]{
        \includegraphics[width=0.45\textwidth]{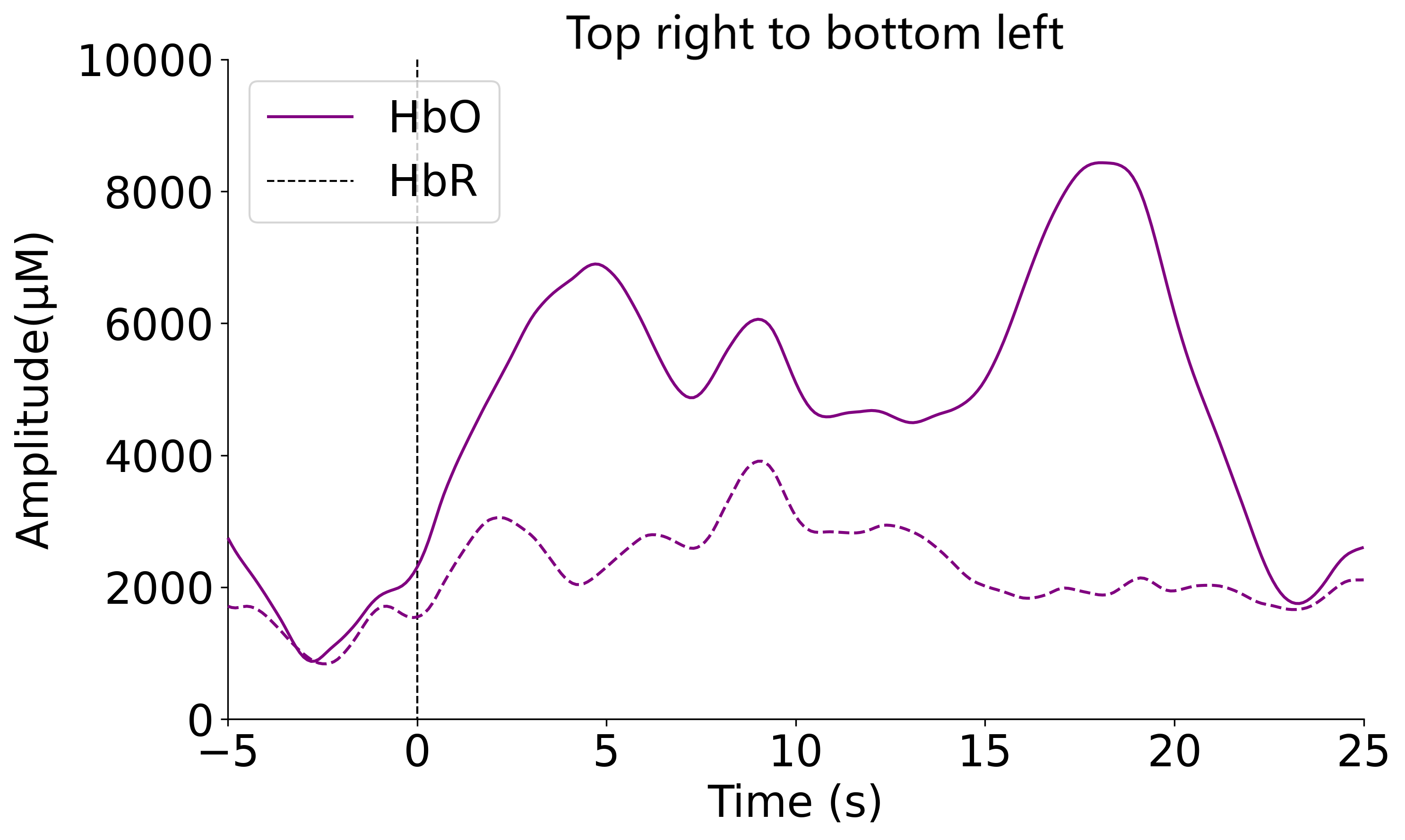}%
      }
    \end{minipage}
  }
  \caption{Average hemodynamic responses (HbO/HbR) for four directional motor imagery tasks.}
  \label{NIRS1}
\end{figure}

\begin{figure}[!h]
	\centering
    \includegraphics[scale=0.55]{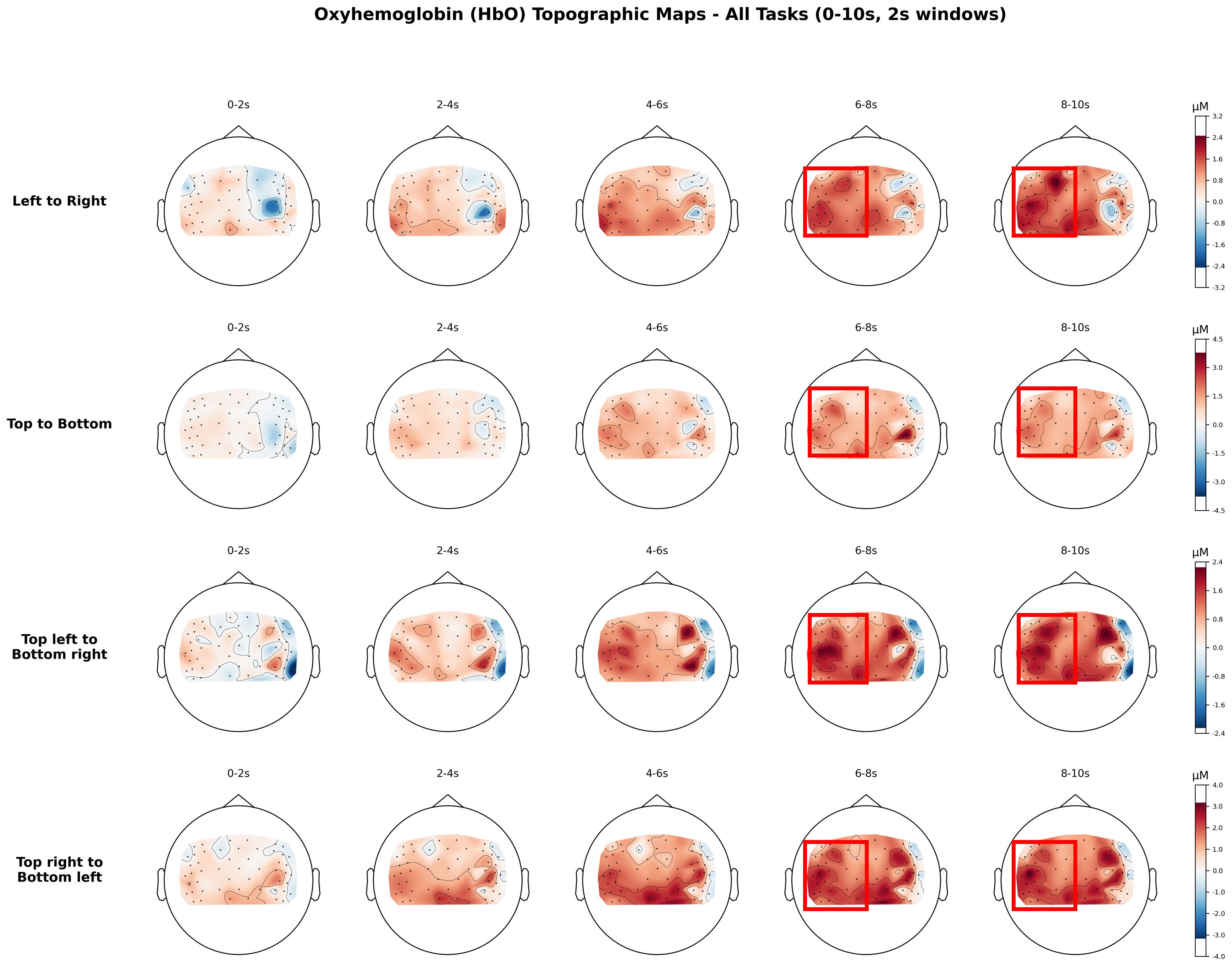}
    \caption{Topographic distribution of HbO responses for four directional motor imagery tasks.}
	\label{Topohbo}
\end{figure}

Fig.~\ref{NIRS1} shows the event-related hemodynamic responses of Subject~1
during the four motor imagery (MI) tasks. Panels~(a)--(d) depict the
averaged time courses of oxygenated hemoglobin (HbO) and deoxygenated
hemoglobin (HbR) for unilateral movements from left to right, top to
bottom, upper-left to lower-right, and upper-right to lower-left,
respectively. The horizontal axis denotes time (seconds), and the vertical axis
denotes concentration ($\mu$M). The results indicate that all four
directional MI tasks evoke a typical hemodynamic response: HbO gradually
increases and reaches a peak around 5--10~seconds after cue onset, and then
slowly returns toward baseline, whereas HbR shows a slight decrease or
remains relatively stable over the same time window. This pattern is
consistent with classical neurovascular coupling, suggesting increased
metabolic demand and local cerebral blood flow in task-related cortical
areas during MI execution. Although the four tasks share similar overall
temporal profiles, they differ in peak amplitude and duration, indicating
that different movement directions within the same upper limb still exhibit
a certain degree of direction-specific modulation.

Fig.~\ref{Topohbo} illustrates the temporal evolution of HbO topographic maps for the
four motor imagery (MI) tasks. From top to bottom, each row corresponds to the
left-to-right, top-to-bottom, top-left-to-bottom-right, and top-right-to-bottom-left tasks, respectively, and from left to right each column represents a different time window from 0–2~seconds to 8–10~seconds after cue onset. Colors indicate changes in
HbO concentration ($\mu$M), with red denoting increases and blue denoting
decreases. The red-outlined regions reach their peak response between 6 and
10~seconds. Across the four tasks, the activated areas show substantial overlap with
similar centers but different boundaries, indicating that HbO in the left
hemisphere increases over time within a largely shared network, while subtle
spatial differences reflect direction-specific modulation.

\subsection*{fNIRS--EEG classification}

\begin{figure}[ht]
  \centering
  \resizebox{1.0\textwidth}{!}{ 
    \begin{minipage}{\textwidth}
      \centering
      \subfloat[EEG Results]{
        \includegraphics[width=0.32\textwidth]{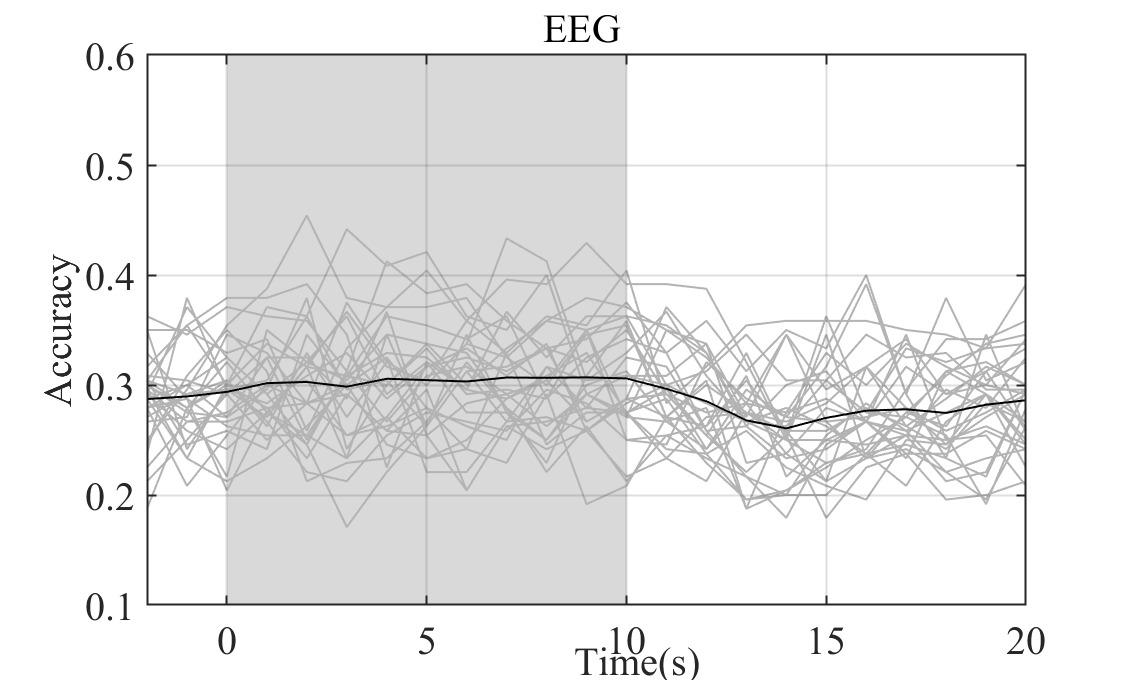}%
      }\hfill
      \subfloat[HbO Results]{
        \includegraphics[width=0.32\textwidth]{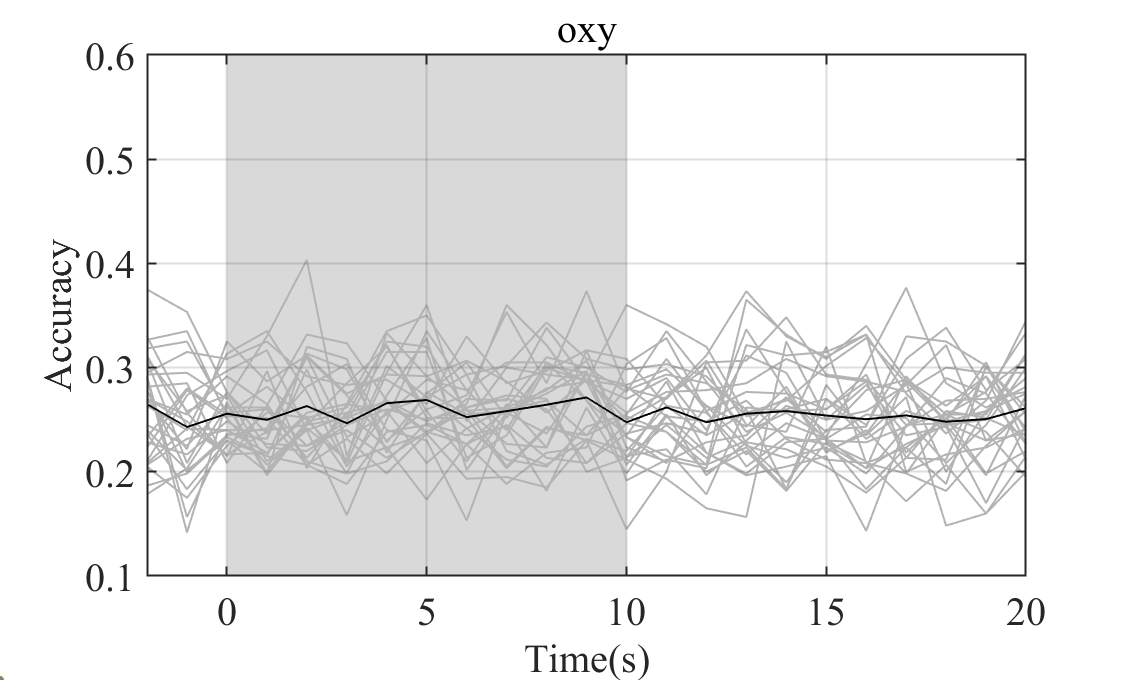}%
      }\hfill
      \subfloat[HbR Results]{
        \includegraphics[width=0.32\textwidth]{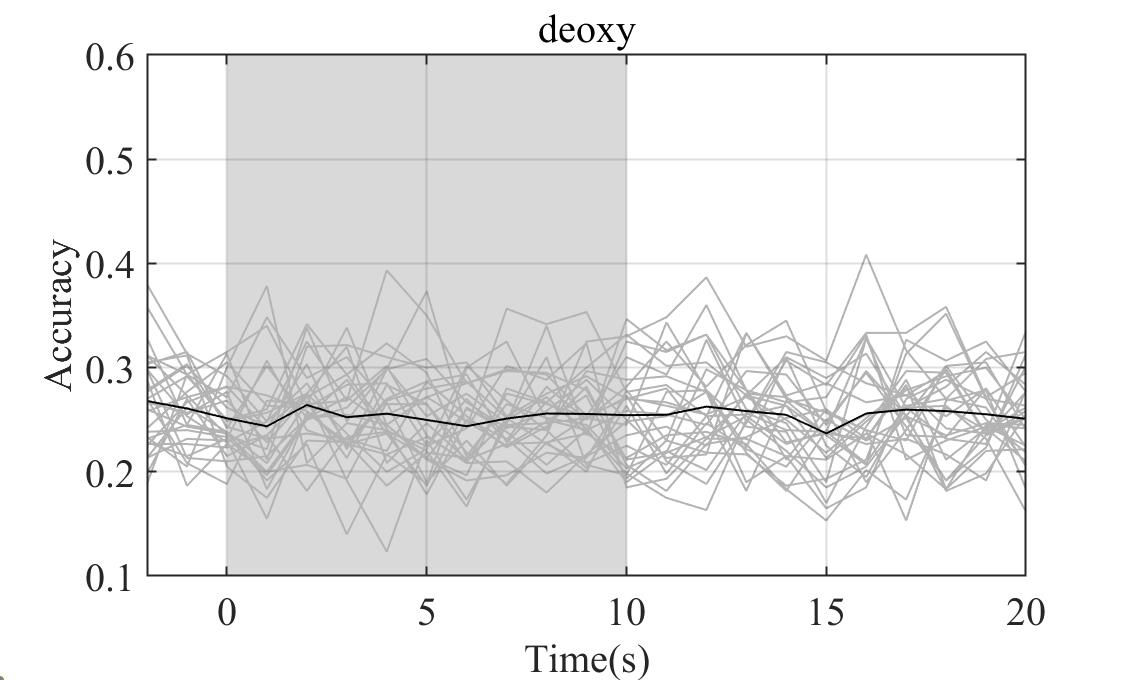}%
      }\\[2mm]
      \subfloat[HbO-HbR Results]{
        \includegraphics[width=0.32\textwidth]{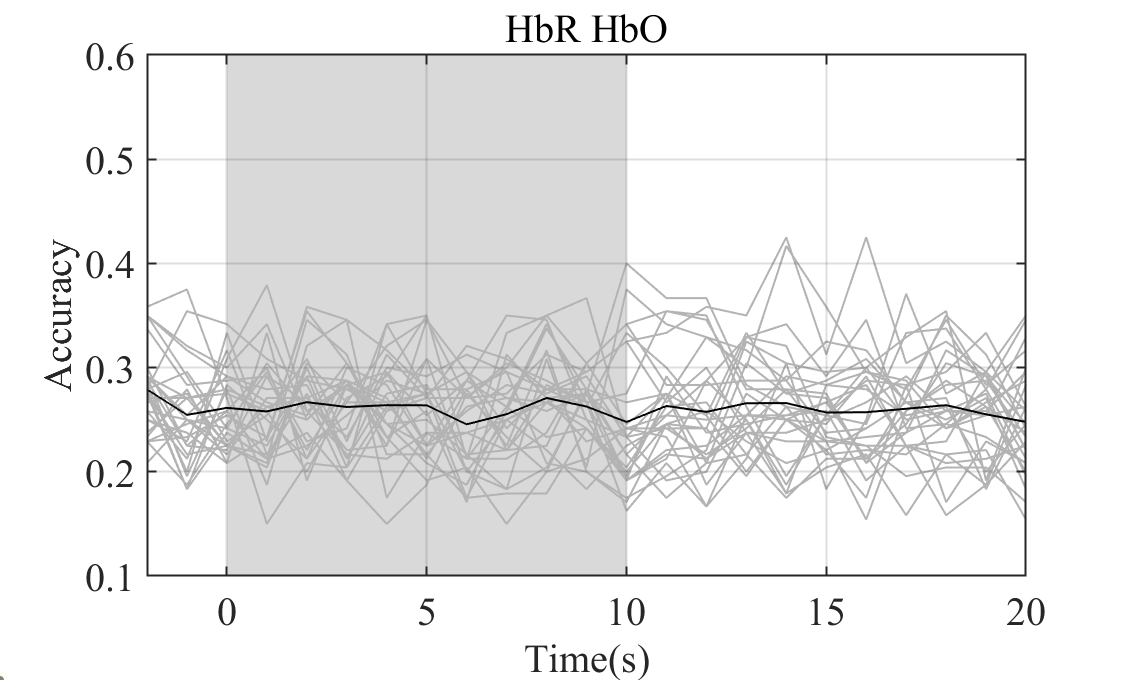}%
      }\hfill
        \subfloat[HbR-EEG Results]{
        \includegraphics[width=0.32\textwidth]{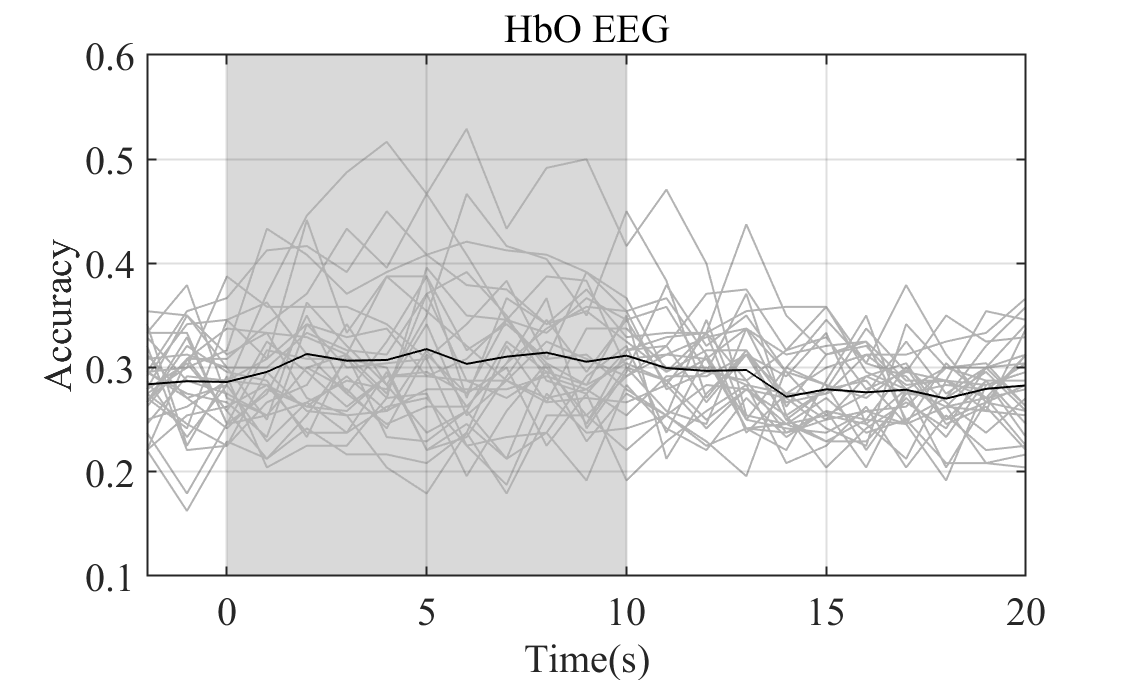}%
      }\hfill
            \subfloat[HbO-EEG Results]{
        \includegraphics[width=0.32\textwidth]{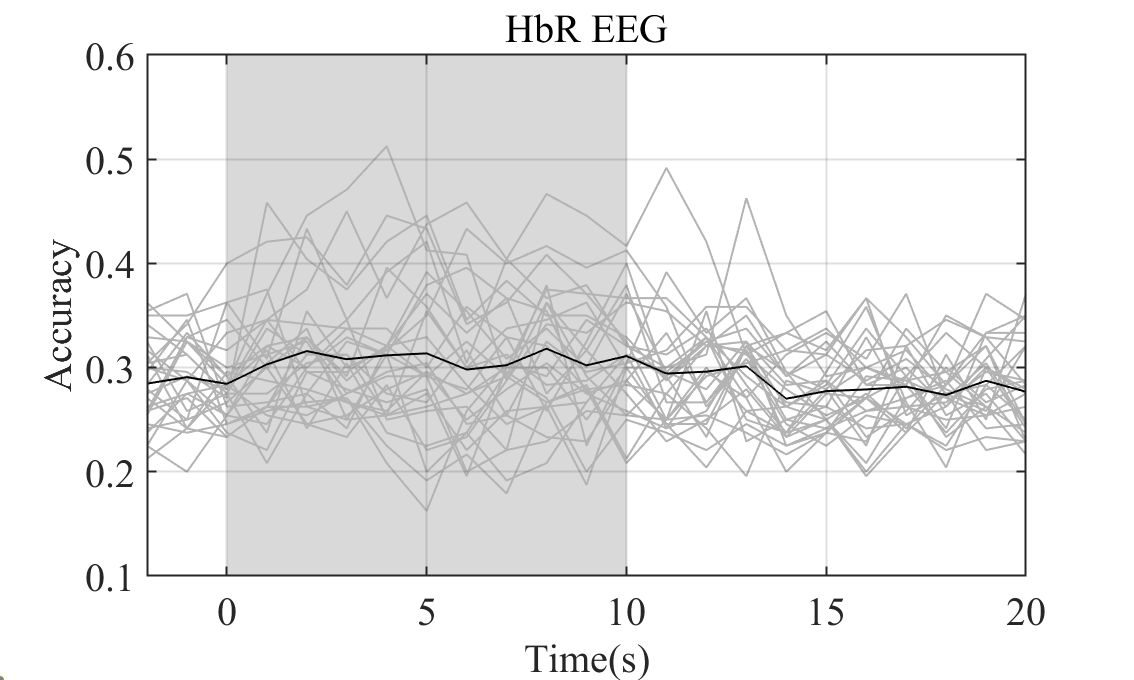}%
      }\\[2mm]     
      \subfloat[HbR-HbO-EEG Results]{
        \includegraphics[width=0.32\textwidth]{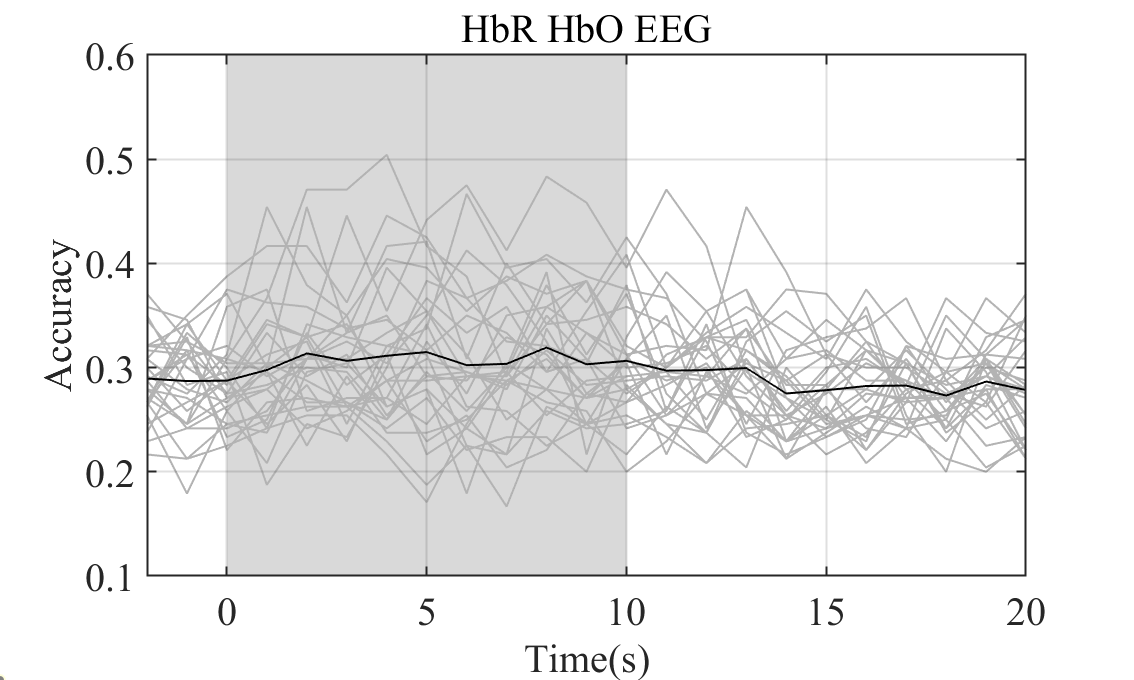}%
      }
    \end{minipage}
  }
  \caption{EEG, fNIRS, and fNIRS+EEG classification accuracies for 3 s moving time window for MI. The x-axis indicates the right edge of the moving time window. Gray lines show the individual classification accuracies while thick black lines show the average ones over whole subjects. The task starts at 0 s and finishes at 10 s. Gray shaded areas indicate task periods.}
  \label{Results}
\end{figure}

All classification analyses are performed in MATLAB R2019b using the BBCI
toolbox~\cite{bbci}. For EEG, features were extracted with a sliding‐window approach
(window length 3~seconds, step size 1~second) starting from $-5$~second and moving to 20~second
relative to cue onset. Thus, the first 3~seconds window covered the interval
$[-5, -2]$~seconds, so that the time axis in the visualization started at $-2$~second.
Filter Bank Common Spatial Pattern (FBCSP) was applied to the band-passed EEG
to obtain spatially filtered signals~\cite{ang2008fbcsp}. FBCSP extends the standard Common
Spatial Pattern (CSP) algorithm by splitting the signal into multiple
frequency bands and performing feature selection, thereby enhancing spatial
discriminability. For each window we computed the log-variance of 28 CSP
components and sorted them in descending order according to their fraction of
the median variance. Compared with eigenvalue-based selection, this
variance-fraction criterion is more robust to outlier trials. The resulting
CSP components captured the most discriminative information within each
sliding window.

Sparse Linear Discriminant Analysis (sLDA) was used as the base classifier.
Because sLDA is inherently binary, we adopted a one-vs-rest strategy: four
independent sLDA classifiers were trained, each separating one class from the
remaining three classes~\cite{shin2017eegfnirs}. The final class label for each window was obtained
by selecting the classifier with the highest discriminant score. To evaluate
performance, we used $10 \times 5$-fold cross-validation and assessed
classification accuracy at every sliding window. Importantly, the spatial
FBCSP filters were estimated only from the training data and then applied to
all windows of the corresponding test data.

For fNIRS, we used the same sliding windows as for EEG and computed two types
of features for each channel: (i) the mean of $\Delta$HbO and $\Delta$HbR, and
(ii) the mean slope of $\Delta$HbO and $\Delta$HbR over the window, yielding
a feature vector with 51 channels and two chromophores per window. The same
sLDA classifier design and cross-validation scheme as for EEG were used for
fNIRS classification~\cite{shin2017eegfnirs}.

To assess the potential benefit of multimodal fusion, we combined the outputs
(LDA projection scores) of the single-modality EEG and fNIRS classifiers into
a meta-classifier feature vector. We examined all combinations
(HbR+HbO, EEG+HbR, EEG+HbO, and EEG+HbR+HbO). The cross-validation procedure
for the meta-classifier was identical to that used for the unimodal EEG and
fNIRS classifiers.

Fig.~\ref{Results} shows the time-resolved four-class MI classification accuracies for
EEG, fNIRS and their fusion using 3~seconds sliding windows. The maximum mean
accuracies over time reached 30.68\% for EEG, 26.40\% for HbR and 27.13\%
for HbO. After modality fusion, the maximum mean accuracies were 27.08\% for
HbR+HbO, 31.78\% for EEG+HbO, 31.81\% for EEG+HbR and 31.92\% for
EEG+HbR+HbO.

\section*{Code Availability}
The usage instructions for this dataset are openly available in our GitHub mirror repository (\url{https://github.com/useflf/Multimodal-fNIRS-EEG-Dataset}). 
The folder $Code/Preprocessing/$ contains preprocessing scripts for both EEG and fNIRS. For EEG preprocessing, $/EEG\_process.py$ and $/EEGPreprocess.m$ are provided. For fNIRS preprocessing, $/2024\_11\_11\_snirf\_trans\_merge.ipynb$ is used for automatic event labeling, and $/no\_mrk.ipynb$ is used for manual label refinement for sessions $0927$, $0928$, $1004$, and $1020$. The folder $Code/Plot/$ includes scripts for EEG time--frequency analysis and topographic mapping, as well as fNIRS averaged hemodynamic response analysis and topographic visualization. The folder $Code/Classification/$ provides unimodal and multimodal classification code based on the FBCSP+sLDA pipeline. All researchers are free to download, use, and cite these resources. We recommend reading the accompanying documentation carefully before using the dataset, and following the original acquisition protocol and ethics statement.





\end{CJK*}

\end{document}